\documentclass[10pt,letterpaper]{article}
\usepackage[left=1 in, top=1 in, right=1 in, bottom=1 in]{geometry}
\usepackage{placeins}

\usepackage{graphicx}

\usepackage{amsmath}
\usepackage{amsfonts}
\usepackage{amssymb}
\usepackage{multicol}
\usepackage{fancyhdr}
\usepackage{bm}

\usepackage{enumitem}
\setlist[itemize]{noitemsep,topsep=0pt}

\usepackage[numbers,sort&compress]{natbib}
\parskip = \baselineskip

\newcommand{\erf}{\mathop{\mathrm{erf}}\nolimits}
\newcommand{\erfc}{\mathop{\mathrm{erfc}}\nolimits}

\begin{document}
\vspace*{0.2in}
\begin{flushleft}
{\Large
\textbf\newline{Chirality provides a direct fitness advantage and facilitates intermixing in cellular aggregates} 
}
\newline
\\
Ashish B. George\textsuperscript{1},
Kirill S. Korolev\textsuperscript{2\dag}
\\
\bigskip
\textbf{1} Department of Physics, Boston University, Boston, MA 02215, USA
\\
\textbf{2} Department of Physics and Graduate Program in Bioinformatics,Boston University, Boston, MA 02215, USA
\\

\bigskip

* ashish.b.george@gmail.com
\dag korolev@bu.edu

\end{flushleft}

\section*{Author summary}
Is it better to be left- or right-handed? The answer depends on whether the goal is making a handshake or winning a boxing match. The need for coordination favors the handedness of the majority, but being different could also provide an advantage. The same rules could apply to microbial colonies and cancer tumors. Like humans, cells often have handedness~(chirality) that reflects the lack of mirror symmetry in their shapes or movement patterns. We find that cells gain a substantial fitness advantage by either increasing the magnitude of their chirality or switching to the opposite handedness. Selection for specific chirality can overcome differences in growth rate and is mediated by the formation of bulges along the colony edge in regions where cells with different chiralities meet.
\section*{Abstract}
Chirality in shape and motility can evolve rapidly in microbes and cancer cells. To determine how chirality affects cell fitness, we developed a model of chiral growth in compact aggregates such as microbial colonies and solid tumors. Our model recapitulates previous experimental findings and shows that mutant cells can invade by increasing their chirality or switching their handedness. The invasion results either in a takeover or stable coexistence between the mutant and the ancestor depending on their relative chirality. For large chiralities, the coexistence is accompanied by strong intermixing between the cells, while spatial segregation occurs otherwise. We show that the competition within the aggregate is mediated by bulges in regions where the cells with different chiralities meet. The two-way coupling between aggregate shape and natural selection is described by the chiral Kardar-Parisi-Zhang equation coupled to the Burgers' equation with multiplicative noise. We solve for the key features of this theory to explain the origin of selection on chirality. Overall, our work suggests that chirality could be an important ecological trait that mediates competition, invasion, and spatial structure in cellular populations.


\newpage
\section*{Introduction}
Living systems have harnessed a variety of physical principles to design and exploit spatial patterns~\cite{murray:mathematical_biology, kondo:rd_biology, whitesides:self_assembly}. Many biological patterns are chiral, i.e. they break left-right symmetry. While the mechanism of chiral symmetry breaking have been elucidated in some systems~\cite{kusmin:parity, quack:parity, blackmond:homochirality_review, bershadsky:chiral_actin, julicher:active_torque_chiral, hejnol:embryonic_evolution, levin:introduction_chirality}, the functional role of chirality remains largely unexplored~\cite{whitton:helix_reversal, hoso:snails, ben_jacob:cooperative_so, levin:introduction_chirality}. 

Chirality exists at all scales: from molecules to populations~\cite{chirality:book, levin:review_chirality, cytoskeletal_chirality:highlight, bershadsky:chiral_actin, ben_jacob:cooperative_so, hallatschek:sectors, korolev:amnat}. The origin of molecular chirality is typically attributed to chance~\cite{frank:chirality, goldenfeld:chirality, kusmin:parity, quack:parity}. For nucleotides and amino acids, the classical explanation of homochirality posits two steps: a fluctuation that slightly breaks the left-right symmetry and a self-amplifying process that increases the asymmetry further~\cite{frank:chirality, blackmond:homochirality_review}. More recent work demonstrates that the amplification step may not be necessary because intrinsic noise in chemical reactions is sufficient to establish and stabilize the symmetry breaking~\cite{goldenfeld:chirality}. The existence of many chiral components within the cell then serves as a natural explanation for macroscopic chirality~\cite{levin:chiral_embryo, matsuno:chiral_development, levin:review_chirality}. Consistent with this view, chiral body plans arise early in the development due to a symmetry breaking event at a microscopic scale, which is amplified further during the subsequent growth~\cite{levin:chiral_embryo, amack:cilia_chiral, hejnol:embryonic_evolution, cytoskeletal_chirality:highlight, julicher:active_torque_chiral, julicher:chirality_mechanism}. Similarly, the macroscopic chirality of bacterial colonies is typically explained by the chirality of individual bacteria~\cite{ben_jacob:cooperative_so, nelson:disordered_review, oddershede:chirality}. 

The existing theory explains how, but not why chirality emerges. Indeed, a lot of effort went into elucidating the mechanism of the chiral symmetry breaking~\cite{chirality:book,  bershadsky:chiral_actin, ben_jacob:cooperative_so, ben_jacob:chiral_modeling}, but the relationship between chirality and fitness has received much less attention~\cite{whitton:helix_reversal, young:shape, hoso:snails, wan:cancer_chiral, ben_jacob:engineering}. Several lines of evidence, however, do suggest that a change in chirality could be advantageous~\cite{whitton:helix_reversal, young:shape, hoso:snails, wan:cancer_chiral, ben_jacob:engineering}. Experiments with \textit{Arthrospira} showed that this bacterium changes from a right-handed to a left-handed helix following the exposure to grazing by a ciliate~\cite{whitton:helix_reversal, young:shape}. Extensive work with~\textit{Paenibacillus} demonstrated that this microbe switches between a chiral and a non-chiral forms to optimize its fitness in different environments~\cite{ben_jacob:cooperative_so, ben_jacob:engineering}. Human cells are also known to form chiral patterns. The handedness of these patterns is the same across all tissue types; except, it is reversed in cancer~\cite{wan:cancer_chiral}. Thus, in a variety of systems, a change in chirality co-occurs with the evolution of higher growth, dispersal, or competitive ability. 

Motivated by these striking examples, we decided to explore whether chirality could be a product of natural selection rather than a historical accident. We asked this question in the context of growing cellular aggregates and found that chirality could directly affect fitness through a pattern-formation mechanism. Our results show that, depending on the growth conditions, it could be advantageous for a cell to increase its chirality or to switch its handedness relative to that of the ancestral population. These dynamics often lead to the coexistence of the left- and right-handed forms, which is a major departure from the classic theories of homochirality~\cite{frank:chirality, blackmond:homochirality_review, goldenfeld:chirality}. Coexisting cell types may enjoy additional benefits of chirality because they develop unique spatial structure that facilitates cross-feeding and other social interactions~\cite{korolev:mutualism, momeni:intermixing, muller:mutualism, menon:diffusion}.


\section*{Results}

\begin{figure}
\begin{center}
\includegraphics[width=8.7cm]{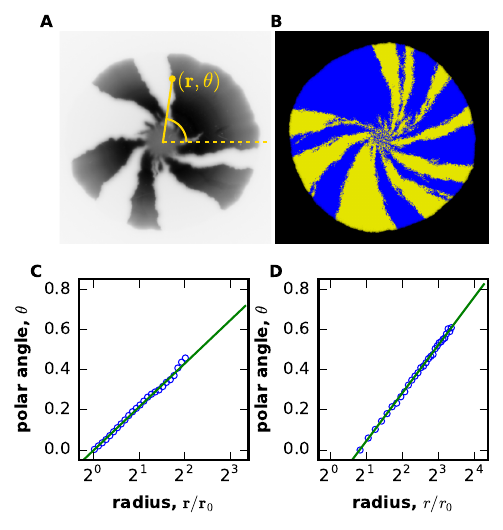}
\caption{\textbf{Reaction-diffusion model of chiral growth accurately describes the behavior of sector boundaries in compact microbial colonies.} Population dynamics are visualized by the spatial pattern produced during the growth of two neutral strains expressing different fluorescent proteins. The growth is largely limited to the colony edge, so the patterns behind the front do not change over time. Although initially the strains are well-mixed, strong genetic drift leads to local extinctions of one of the strains, which manifests as a characteristic pattern of sectors in both experiments~\textbf{(A)} and simulations~\textbf{(B)}. The boundaries between the sectors fluctuate due to genetic drift and twist counterclockwise due to a chiral bias in cell motion. This bias is quantified in~\textbf{(C)} for experiments and in~\textbf{(D)} for simulations by plotting the polar angle~$\theta$, averaged over many sector boundaries,~\textit{vs.} the radius~$r$. A constant boundary velocity along the colony edge should result in a linear increase of~$\theta$ with~$\ln r$~\cite{korolev:sectors}. Consistent with this expectation, both plots show that sector boundaries are logarithmic spirals. The excellent agreement between experiments and simulations indicates that our reaction-diffusion model is suitable for the study of competition between chiral strains in compact microbial colonies. The experimental data was obtained from the Dryad digital data repository associated with Ref.~\cite{korolev:amnat}. Here,~$m_0=m_s=m_b=m_d=0$,~$g=0.03$,~$N=100$,~$m_l=0.045$,~$m_r=0.005$ for both strains. Radius of initial circle was~$30$ on a lattice of~$700 \textsf{x} 700$ sites.
}
\label{fig:model_works}
\end{center}  
\end{figure}

\subsection*{Model of chiral growth in compact aggregates}
Microbial colonies and cancer tumors exhibit a variety of complex morphologies including smooth and rough compact disks, concentric rings, radiating branches, and many others~\cite{berg:chemotaxis_patterns_1, shapiro:scientific_american, matsushita:diagram, ben_jacob:cooperative_so, anderson:tumor_morphology}. For aggregates that grow as a network of filaments or branches, chirality manifests as clockwise or counterclockwise bending of the branches~\cite{ben_jacob:chiral_modeling}. Although chiral growth is not as easily detected for other morphologies, it could be present even if the overall colony shape shows no left-right asymmetry.

The ``hidden'' chirality can be revealed by growing a colony from an initially well-mixed population of two strains that are identical, except they express different fluorescent proteins. As the colony expands, demographic fluctuations at the colony edge lead to local extinctions of one of the strains creating a characteristic pattern of sectors shown in Fig.~\ref{fig:model_works}A~\cite{hallatschek:sectors, korolev:amnat, korolev:sectors, freese:conjugation, korolev:review}. In the absence of chirality, the boundaries between the sectors are approximately radial, but the boundaries twist consistently clockwise or counterclockwise for chiral cells. The direction of this twisting stays the same across replicate colonies~\cite{korolev:amnat, oddershede:chirality, hallatschek:sectors}. So far, this method has been applied only to a few model organisms; two of them, \textit{Escherichia coli} and \textit{Bacillus subtilis}, were found to grow in a chiral fashion~\cite{korolev:amnat, oddershede:chirality, hallatschek:sectors}. Since most organisms have not been examined, hidden chirality could potentially be quite prevalent in cellular aggregates.

The twisting of the boundaries can be quantified by the increase of the polar angle with the distance from the colony center. Using the data from Ref.~\cite{korolev:amnat}, Figure~\ref{fig:model_works}C shows that this dependence is logarithmic, i.e. the boundaries twist as Bernoulli spirals~\cite{huntley:divine}. The origin of the logarithmic twisting can be explained by a simple phenomenological description that combines a constant velocity of the sector boundary with a linear increase of the colony radius in time~\cite{korolev:sectors}. The molecular mechanism responsible for cellular chirality has not been fully determined, but we do know that chirality in not due to flagella and is mediated by outer membrane proteins such as antigen 43, extracellular structures including pili, and the interaction with the substratum~\cite{oddershede:chirality}.

Because the factors that mediate chirality also contribute to other components of cell phenotype, it could be challenging to create two strains that differ only in their chirality. This difficulty, however, can be easily overcome in a computational model, where chirality can be tuned without affecting the growth and motility~(see Methods and SI). A large number of approaches has been developed to model cellular aggregates from analytic equations to mechanistic simulations~\cite{korolev:wave_splitting, xavier:framework, cell_modeller, gro, waclaw:mechanical}. We chose to study a minimal reaction-diffusion model because it involves few parameters and is more likely to capture the universal behavior that generalizes across diverse cellular populations. We also focused on the simplest morphology of a compact disk because it is both common and well-understood.

For strains with equal chirality, our simulations~(described below) showed excellent agreement with the experimental observations (Fig.~\ref{fig:model_works}). The simulations not only reproduced the formation and bending of sectors, but also exhibited the same logarithmic twisting of sector boundaries as in the experiments. Thus, the few ingredients in our model are sufficient to describe the chiral growth in compact microbial colonies.

In simulations, cells grow and move in a two-dimensional habitat. The movement is stochastic and short-ranged, but potentially biased relative to the direction of the local density gradient. For non-chiral populations, the bias is along the gradient, in the direction of the outward growth. This bias accounts for the effects of chemotaxis towards nutrients and higher pressure within the colony. For chiral cells, the direction of movement is not collinear with the applied force~\cite{bechinger:active_crowded}, so the bias in cell movement makes a nontrivial angle with the local gradient of the population density. The sign and magnitude of this angle control the handedness and the strength of chirality. A detailed description of the simulation procedure is provided in Methods and SI.

In the deterministic limit, our simulations are described by the following reaction-diffusion equation: 

\begin{equation}
\label{model}
\frac{\partial n^{(\alpha)}}{\partial t} = g^{(\alpha)}n^{(\alpha)} - \boldsymbol{\nabla}\boldsymbol{\cdot}\boldsymbol{J}^{(\alpha)},
\end{equation}

\noindent where~$t$ is time,~$n^{(\alpha)}$ is the population density of strain~$\alpha$,~$g^{(\alpha)}$ is a density-dependent per capita growth rate, and the flux~$J$ is given by

\begin{equation}
\boldsymbol{J}^{(\alpha)}_{i} = -D^{(\alpha)} \boldsymbol{\nabla}_{i}n^{(\alpha)} - n^{(\alpha)}\sum_{j}\left(S^{(\alpha)}\delta_{ij}- A^{(\alpha)} \epsilon_{ij}\right)\boldsymbol{\nabla}_{j}n.
\label{flux}
\end{equation}

\noindent Here, the indexes denote the Cartesian components of vectors,~$\delta_{ij}$ is the unit tensor, and~$\epsilon_{ij}$ is the totally antisymmetric tensor, also known as Levi-Civita symbol.

The density-dependent diffusion and advection are described by~$D^{(\alpha)}(n)$ and~$S^{(\alpha)}(n)$ respectively. $A^{(\alpha)}(n)$ is the strength of the chiral term, which is the only term that changes sign under the mirror symmetry. To the lowest order in the gradient expansion, no other term that breaks the left-right symmetry can be added to Eq.~(\ref{flux}), which suggests that all chiral patterns in compact aggregates are described by our coarse-grained theory regardless of the microscopic origin of chirality. 

\subsection*{Competition between cells with different chirality}
\begin{figure}[!htb]
\begin{center}
\includegraphics[width=8.7cm]{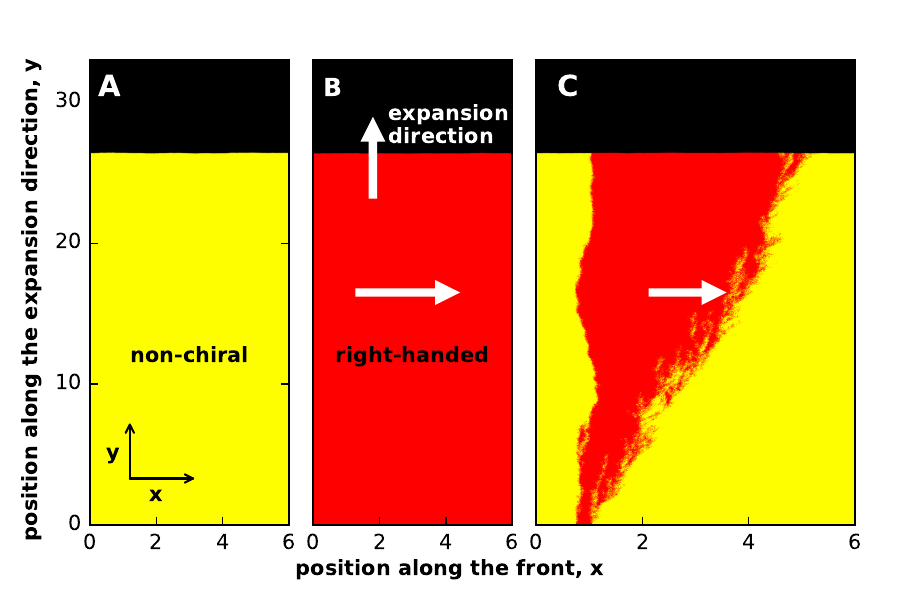}
\caption{\textbf{Chirality provides a fitness advantage in competition, but not in overall growth.} The first two panels show that a non-chiral~\textbf{(A)} and a chiral strain~\textbf{(B)} expand with the same velocity. To facilitate this comparison, we started the simulations with a linear instead of a circular front. Panel~\textbf{(C)} demonstrates the chiral strain displaces the non-chiral strain when they compete within the same colony: A small initial population of the right-handed strain~(shown in red) expands over time and eventually takes over. Because the simulations respect the mirror symmetry, the results were the same irrespective of whether the chirality was due to a left-handed or a right-handed bias in motility. Note that the fate of a strain is not determined solely by its expansion velocity because the colonies expand as pushed waves~\cite{saarloos:review, gandhi:pulled_pushed, birzu:semi_pushed, waclaw:mechanical}. The expansion velocity of a pushed wave depends not only on the growth and migration rates at the leading edge, but also on the non-linear population dynamics within the wave front. As a result, the outcome of the competition is affected by how each of the strains responds to the presence of the other strain. In our simulations, expansions are pushed because of the density-dependent motility; see Methods and SI. Here,~$m_0=m_s=m_b=m_d=0$,~$g=0.1$,~$N=200$ for both strains. $m_l^{(1)}=m_r^{(1)}=0.05$,~$m_l^{(2)}=0.01$,~$m_r^{(2)}=0.09$ on a lattice of~$600 \textsf{x} 3300$ sites. All distances were measured in~$100$ units where~$\Delta \textsf{x}=\Delta \textsf{y}=1$ unit.}
\label{fig:chiral_wins}
\end{center}  
\end{figure}

To test whether chirality confers a selective advantage, we competed a chiral strain \textit{vs}. a non-chiral strain. The growth and motility rates of these strains were identical and, as a result, they expanded at the same rate when grown separately~(Fig.~\ref{fig:chiral_wins}AB); see SI for a more quantitative comparison of the velocities. The chiral strain, however, had a clear selective advantage when the competition occurred within the same colony. Figure~\ref{fig:chiral_wins}C illustrates this by showing how the chiral strain took over the population starting from a small, localized patch representing a mutation or an immigration event. This selective advantage of chirality was not specific to the competition between a chiral \textit{vs.} strictly non-chiral strains. In simulations, we explored a wide range of microscopic parameters and invariable found that the more chiral strain outcompeted the less chiral strain \textit{when both strains had the same handedness}.

\begin{figure*}[!htb]
\begin{center}
\includegraphics[width=\linewidth]{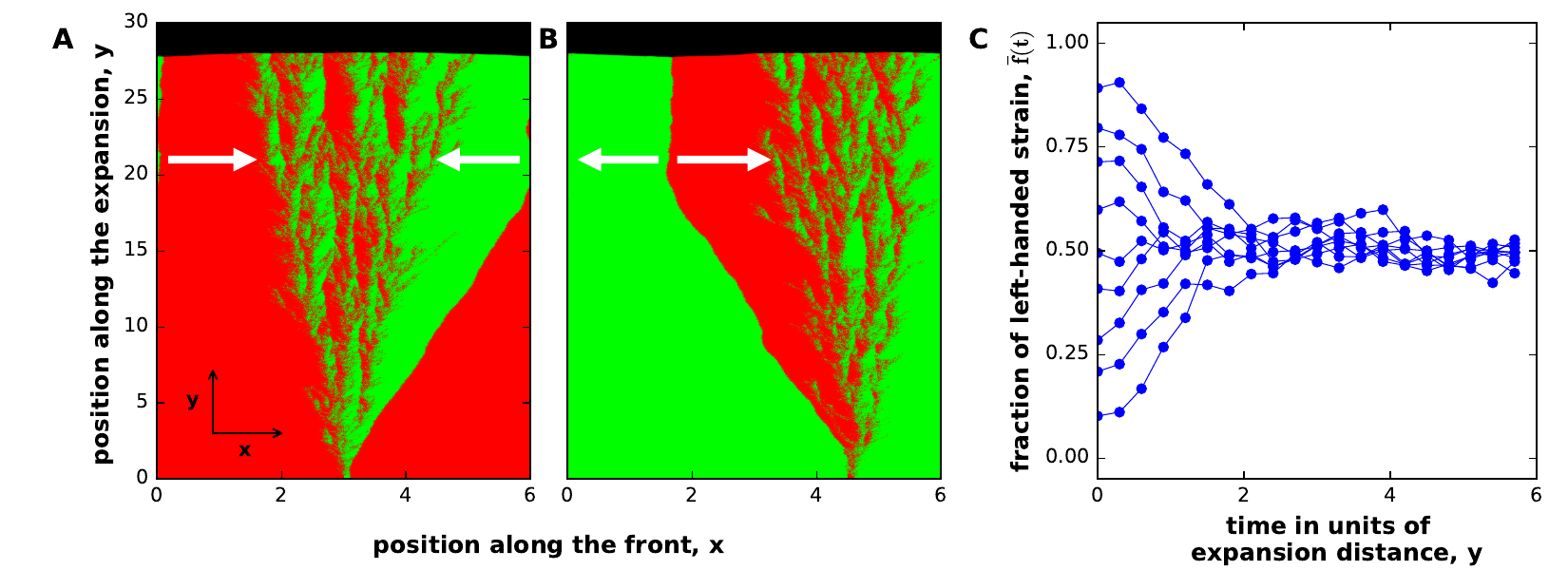}
\caption{\textbf{Selection for coexistence between strains with opposite handedness.} Panel~\textbf{(A)} shows that a left-handed mutant~(shown in green) can invade a right-handed population~(shown in red). The reverse invasion also occurs and is shown in panel~\textbf{(B)}. This negative frequency-dependent selection is further illustrated in panel~\textbf{(C)}, which shows how~$\bar{f}$, the spatially averaged relative abundance of the first strain, changes over time starting from different initial conditions. At~$t=0$, the strains are spatially separated in~(A) and~(B), but well-mixed in~(C). In this figure, the strains have exactly opposite chiralities, but coexistence occurs more generally; see Fig.~\protect{\ref{fig:relative_chirality}}. Note that the selection for coexistence relies on the presence of boundaries between the strains. When strains intermix~(as shown in this figure), we observe a strong and time-invariant selection for coexistence. When strains do not intermix~(see Figs.~\protect{\ref{fig:bulges_boundaries}} and~\protect{\ref{fig:selection}}), the number of boundaries slowly declines over time due to neutral coarsening~\protect{\cite{korolev:review, korolev:amnat}}. In such cases, robust coexistence relies on occasional external re-mixing events, e.g. during the establishment of a new colony. Here,~$m_0=m_s=m_b=m_d=0$,~$g=0.1$,~$N=200$ for both strains. $m_l^{(2)}=0.09$,~$m_r^{(2)}=0.01$,~$m_l^{(2)}=0.01$,~$m_r^{(2)}=0.09$ on a lattice of~$600\textsf{x}3000$ sites.  All distances were measured in~$100$ units where~$\Delta \textsf{x}=\Delta \textsf{y}=1$ unit.}
\label{fig:coexistence}
\end{center}
\end{figure*}

In contrast, the competition between two strains with \textit{opposite handedness} often resulted in stable coexistence. As an example of this, Fig.~\ref{fig:coexistence} shows the competition between two strains with chiralities that equal in magnitude, but opposite in direction. Both strains invaded when introduced in the population of the opposite handedness, but did not completely take over~(Fig.~\ref{fig:coexistence}AB). Instead, the population approached a steady state where both strains were equally abundant. To confirm this observation of negative frequency-dependent selection, we performed simulations starting from well-mixed initial conditions with different fractions of the left-handed strain. As expected from symmetry, the fraction of the left-handed strain converged to~50\%~(Fig.~\ref{fig:coexistence}C) suggesting that left- and right-handed strains can stably coexist. Strains with opposite, but not exactly equal chirality were also found to coexist, but the equilibrium fractions deviated from the~50:50 ratio in favor of the more chiral strain.

\begin{figure*}[!htb]
\begin{center}
\includegraphics[width=15cm]{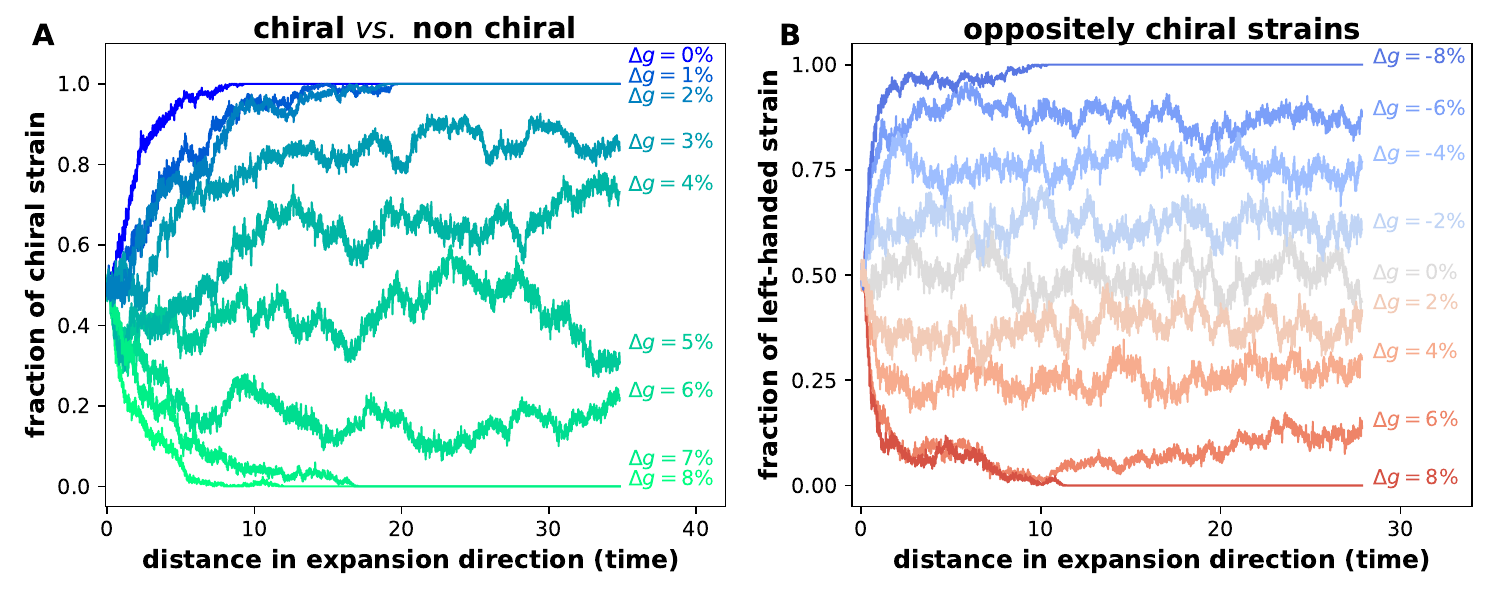}
\caption{\textbf{Effects of chirality persist despite growth rate differences.} We competed strains with different growth rates and chirality starting from well-mixed initial conditions. The abundances of the strains were equal at the beginning of the simulations, but changed over time leading either to stable coexistence or to the exclusion of one of the strains. The outcome of the competition depends on the relative growth rates of the strains quantified by~$\Delta g$. The competition between a left-handed strain and a faster growing non-chiral strain is shown in~\textbf{(A)}. The non-chiral strain is outcompeted even if it has a growth advantage as high as to~2\%. The chiral strain becomes extinct only when its growth penalty exceeds~7\%. For intermediate~$\Delta g$, the two strains stably coexist. Similar dynamics occur during the competition between the strains with equal, but opposite chirality shown in panel \textbf{(B)}. The stable coexistence between the strains is destroyed only by growth rate differences higher than about~7\%. Here,~$m_0=m_s=m_b=m_d=0$ for both strains and~$N=200$. In~(A),~$m_l^{(1)}=0.1$,~$m_r^{(1)}=0.0$, $m_l^{(2)}=0.05$, $m_r^{(2)}=0.05$ on a lattice of~$1000\textsf{x}3600$ sites.We fixed~$g=0.01$ for the left handed-strain and varied the growth rate of the non-chiral strain according to~$g(1+\Delta g/100\%)$. In~(B),~$m_l^{(1)}=0.09$,~$m_r^{(1)}=0.01$, $m_l^{(2)}=0.01$, $m_r^{(2)}=0.09$ on a lattice of~$500\textsf{x}3000$ sites. We set~$g=0.1$ for the left-handed strain, and varied the growth rate of the right-handed strain according to~$g(1+\Delta g/100\%)$. The distances on the x-axis were measured in~$100$ units where~$\Delta \textsf{x}=\Delta \textsf{y}=1$ unit. We verified stable coexistence by starting the simulation above and below the observed steady-state relative abundances and checking that they return to the same steady-state values.}
\label{fig:growth_difference}
\end{center}
\end{figure*}

The effects of chirality persist even when the strains have different growth rates. To demonstrate this, we competed a chiral \textit{vs.} a faster growing non-chiral strain~(Fig.~\ref{fig:growth_difference}A). The chiral strain completely excluded the non-chiral strain provided its growth rate penalty was less than~2\%. For growth rate differences between 2\% and 7\%, the two strains stably coexisted. The chiral strain went extinct only when its growth rate penalty exceeded 7\%. Similar dynamics occurred during the competition between two strains with opposite handedness~(Fig.~\ref{fig:growth_difference}B). As we increased the difference in the growth rate between the strains, their steady-state abundance started to deviate from the~50:50 ratio. The coexistence was lost only when the growth rate difference exceeded 7\%. The 7\% threshold is representative for the parameters used in our reaction-diffusion model, but its value in an actual biological system could be affected by the details of the cell-cell and cell-surface interactions, which we did not explicitly model. Nevertheless, our results demonstrate that chirality can influence the outcome of the competition between strains with substantial differences in growth and motility.

Selection mediated by chirality may seem quite surprising. In the following, we first explain the origin of this phenomenon using the competition between strains with opposite handedness as a simple example. After that, we return to the general case and explore the transition between exclusion and coexistence as the chirality of the strains is varied. We also describe the role of demographic fluctuations and characterize the spatial patterns that emerge in populations of chiral cells.

\subsection*{Effective theory of chiral growth}
Why does chirality affect competition? To answer this question, we developed an analytical theory that explains the spatial patterns shown in Figs.~\ref{fig:model_works}-\ref{fig:coexistence}. For this purpose, we reduced the reaction-diffusion model~(Eq.~(\ref{model})) to a simpler effective theory that describes only the overall shape of the colony edge and its genetic composition~(see SI). This effective theory can also be derived purely from the symmetry considerations and is therefore more general than the underlying reaction-diffusion model~(see SI). Below, we use the effective theory to explain how the competition between two strains is affected by their chiralities. Our main result is that the difference in chiralities leads to changes in colony shape, which in turn influence the relative abundance of the strains. 

Because little growth occurs inside cellular aggregates~\cite{waclaw:mechanical, nadell:sociobiology}, their population dynamics can be largely described in terms of only two variables: the position of expansion front and the relative abundance of the strains at the edge of the colony. To fix the coordinate system, we take the~$x$-axis to be a straight line along the average direction of the colony edge~(Fig.~\ref{fig:chiral_wins}A). The~$y$-axis then points in the direction of colony growth. We denote the~$y$-coordinate of the expansion front by~$h(t,x)$, and the fraction of the first strain by~$f(t,x)$. In terms of these quantities, the effective theory is given by the following set of equations:
 
\begin{equation}
\label{effective_theory}
\begin{aligned}
&\frac{\partial h}{\partial t} = v_0 + \frac{v_0}{2}\left(\frac{\partial h}{\partial x}\right)^2 + D_h \frac{\partial^2 h}{\partial x^2} + \alpha\frac{\partial f}{\partial x} + \mathrm{noise},\\
&\frac{\partial f }{\partial t}  = D_f\frac{\partial^2 f}{\partial x^2} + \beta (f^* - f) \frac{\partial f}{\partial x} +  v_0 \frac{\partial h}{\partial x}\frac{\partial f}{\partial x} + \mathrm{noise}.
\end{aligned}
\end{equation}

The first equation in Eq.~(\ref{effective_theory})  is an extension of the Kardar-Parisi-Zhang~(KPZ) equation of surface growth~\cite{kpz, spohn:exact_kpz, kardar_drossel:KPZ_DP2_binary_alloys}. Here, the first term,~$v_0$, is the expansion velocity of the strains grown in isolation. The second term accounts for the fact that the expansion of a tilted front~($\frac{\partial h}{\partial x}\ne0$) occurs perpendicular to the front and, therefore, at a angle with the~$y$-axis. The third term arises because fronts that are curved outward expand more slowly and because of effective surface tension at the edge of the microbial colony~\cite{giverso:surface_tension_KPZ}. The last term couples the dynamics of~$f$ and~$h$ and is a new term that describes chirality. Its magnitude is controlled by parameter~$\alpha$, which is proportional to the difference in the chiralities of the strains. We show below that this last term changes colony shape and mediates the competition between chiral strains.

The second equation in Eq.~(\ref{effective_theory}) is an extension of the Burgers' equation used to describe fluid and traffic flow~\cite{bateman:burgers_first, burgers:original, sachdev:diffusive_waves, whitham:waves}. The first term describes random, diffusion-like movement, while the second term accounts for the directional motion due to a chiral bias in motility. Here, the factor of~$-\beta(f^*-f)$ can be viewed as a local advection velocity, which depends on the relative abundance of the strains and two parameters that describe the chiral properties of the strains. The first parameter,~$\beta$, is proportional to the difference in the chiralities of the strains. The second parameter~$f^*$ is the ratio of the chirality of the first strain to the difference in the chiralities of the strains. 

We choose the most left-handed~(also the least right-handed) strain to be the first strain, which ensures that~$\alpha$ and~$\beta$ are positive~(see SI). With this convention, the first strain that is left-handed for~$f^*>0$ and right-handed for~$f^*<0$. When~$f^*\in(0,1)$, the two strains have opposite handedness, while, when~$f^*\in(-\infty,0)\cup(1,+\infty)$, the two strains have the same handedness. Furthermore,~$f^*=0$ corresponds to a non-chiral and a right-handed strain, and~$f^*=1$ corresponds to a left-handed and a non-chiral strain. The special case of equal, but opposite chiralities corresponds to~$f^*=1/2$.

The third term in the equation for~$\frac{\partial f}{\partial t}$ describes how colony shape affects the relative abundance of the strains. This term is non-zero only in tilted regions of the front, where cells at higher~$h$ displace cells at lower~$h$ as the colony grows. In other words, the relative abundance of the strains changes because the growth of the colony proceeds in the direction perpendicular to the front and, therefore, induces the movement of cells along the~$x$-axis whenever~$\frac{\partial h}{\partial x}\ne0$. Below we demonstrate how this coupling between colony shape~$h(x)$ and genetic composition~$f(x)$ mediates the competition between the strains in compact aggregates.

Finally, the noise terms account for demographic fluctuations and genetic drift. In the first equation, the noise is the regular additive noise present in the KPZ equation; it arises due to local fluctuations in the growth velocity. In the second equation, the noise accounts for genetic drift, so it is multiplicative with the strength proportional to~$\sqrt{f(1-f)}$. Such dependence on~$f$ is typical for population dynamics~\cite{korolev:review} and is necessary to ensure that~$f=0$ and~$f=1$ are absorbing states.

\subsection*{Bulges and dips at sector boundaries}
Genetic drift in the equation for~$f$ leads to local extinctions of one of the strains and the formation of sector boundaries~\cite{korolev:review}; see Fig.~\ref{fig:model_works}. When these boundaries separate strains with different chirality,~$h(t,x)$ develops this characteristic shape that ultimately controls the competition between the strains. For simplicity, we first discuss the behavior of two strains with exactly opposite chiralities~($f^*=1/2$). In this special case, the mirror symmetry ensures that the boundaries between the strains do not have a net bias and, therefore, remain stationary when~$\frac{\partial h}{\partial x}=0$.

\begin{figure*}
\begin{center}
\includegraphics[width=\linewidth]{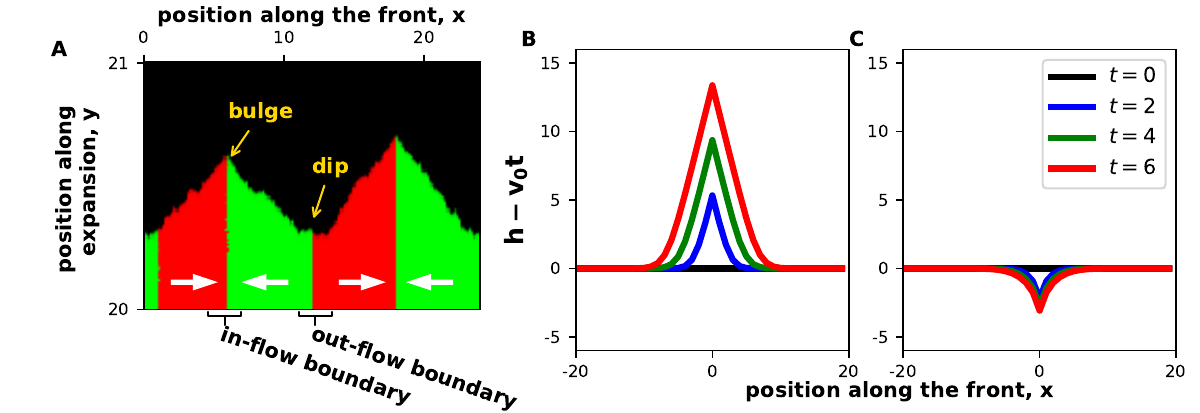}
\caption{\textbf{Boundaries between strains with different chiralities create front undulations.} \textbf{(A)} A magnified view of the colony front shows bulges and dips near in-flow and out-flow boundaries. The analytical solutions for the shape of bulges and dips are shown in panels~\textbf{(B)} and~\textbf{(C)} respectively. Note that both the theory and the simulations predict an approximately triangular bulge shape. Here,~$m_0=m_s=m_b=m_d=0$,~$g=0.1$,~$N=200$ for both strains, and~$m_l^{(1)}=0.009$, $m_r^{(1)}=0.001$, $m_l^{(2)}=0.001$, $m_r^{(2)}=0.009$ on a lattice of~$2400\textsf{x}2100$ sites in panel(A).  The chirality of the strains is shown with thick white arrows. All distances were measured in~$100$ units where~$\Delta \textsf{x}=\Delta \textsf{y}=1$ unit.}
\label{fig:bulges_boundaries}
\end{center} 
\end{figure*}

The dynamics of a strain boundary depends on whether the chiral biases of the strains point towards or away from it. We term a boundary an in-flow boundary when the strains move towards each other, i.e. the left-handed strain is to the right of the boundary, and the right-handed strain is to the left of the boundary~(Fig.~\ref{fig:bulges_boundaries}). Boundaries with the opposite arrangement of the strains are termed out-flow boundaries.

Equation~(\ref{effective_theory}) predicts that the two types of boundaries have a diametrically opposite effect on the colony shape. For in-flow boundaries~$\alpha\frac{\partial f}{\partial x}>0$, and we expect a bulge due to local overgrowth of~$h$. In contrast, a dip in the front is expected at out-flow boundaries, where~$\alpha\frac{\partial f}{\partial x}<0$. Figure~\ref{fig:bulges_boundaries} shows that these shapes indeed develop in our simulations.  

We exactly solved the chiral KPZ equation without noise and obtained an analytical expression for the shapes of the bulges and dips~(see SI) in the limit of sharp boundaries between the strains. After a transient, the bulges assume an approximately triangular shape given by

\begin{equation}
\label{bulge_shape}
h(t,x)-v_0t = \left\{
		\begin{aligned}
		& 0,\;\;|x-x_b|\ge\frac{v_0\alpha}{4D_h}t,\\
		& \frac{v_0\alpha^2}{8D_h^2}t-\frac{\alpha}{2D_h}|x-x_b|,\;\;|x-x_b|<\frac{v_0\alpha}{4D_h}t\\
		\end{aligned} 
		\right.
\end{equation}

\noindent for an in-flow boundary located at~$x_b$. The slope of the bulge stays constant, but its height and width increase linearly in time. The depth of the dips, on the other hand, increases only logarithmically in time, so they remain quite small on the time scale of our simulations~(see SI). As a consequence, the front primarily consists of titled regions near the bulges and flat regions away from the strain boundaries~(Fig.~\ref{fig:bulges_boundaries}).

One immediate consequence of Eq.~(\ref{bulge_shape}) is that a mixture of two strains with opposite handedness expands faster than either strain in isolation. Indeed, $dh(t,x_b)/dt= v_0 +v_0\alpha^2/(8D_h^2)$, which is greater than~$v_0$. In our simulation, this increase was typically on the order of a few percent; see SI for further details. 

\subsection*{Origin of selection}
\begin{figure*}
\begin{center}
\includegraphics[width=\linewidth]{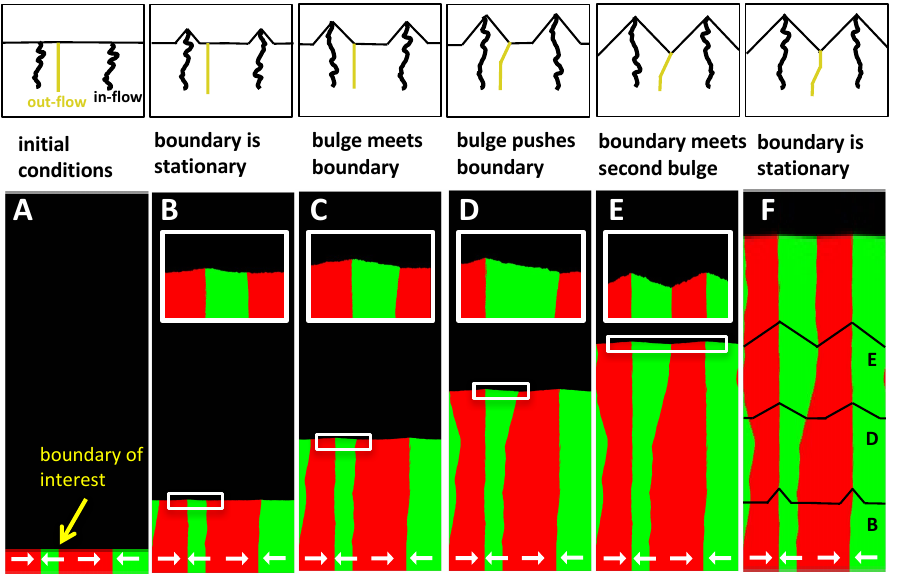}
\caption{\textbf{Bulges drive selection by ``pushing'' out-flow boundaries.} The panels show the motion of an out-flow boundary due to the expansion of the bulges formed at the surrounding in-flow boundaries. Initially, the bulges are small~\textbf{(AB)}. Bulge growth has no effect, until one of the bulges comes in contact with the boundary of interest~\textbf{(C)}. After that, the expansion of the bulge displaces the out-flow boundary~\textbf{(D)}. The movement of the boundary stops when it is locked between two nearest bulges~\textbf{(EF)}. The relationship between the location of bulge and the motion of the out-flow boundary is further clarified in schematics above each of the panels. The black lines in the last panel show the locations~(but not the actual size) of the bulges at earlier times. To illustrate the dynamics most clearly, we chose the model parameters in the no-mixing regime~(see Fig.~\protect{\ref{fig:mixing}}). As a result, both in-flow and out-flow boundaries appear almost equally sharp. The chirality of the strains is shown with thick white arrows. Here,~$m_0=m_s=m_b=m_d=0$,~$g=0.1$,~$N=200$ for both strains. $m_l^{(2)}=0.009$,~$m_r^{(2)}=0.001$,~$m_l^{(2)}=0.001$,~$m_r^{(2)}=0.009$ on a lattice of~$2400\textsf{x}6400$ sites.}
\label{fig:selection}
\end{center}  
\end{figure*}

The expansion of bulges changes the relative abundance of the strains~(Fig.~\ref{fig:selection}). Initially, no bulges are present because we start our simulations with a flat front to mimic the coffee-ring effect that creates a smooth edge around a microbial colony~\cite{yunker:coffee_ring}. As the colony expands, small bulges form and grow around the in-flow boundaries. In the beginning, a small bulge has no effect on the genetic composition of the front because it is completely enclosed within the two sectors surrounding the in-flow boundary. However, as the bulge expands, it comes in contact with the out-flow boundaries on its sides and then starts ``pushing'' them outwards. The subsequent movement of the out-flow boundaries changes sector sizes and, therefore, the relative abundance of the strains. 

For the case of~$f^*=1/2$ that we are considering now, the left and the right ends of the bulge are equidistant from the in-flow boundary. Hence, the two strains have equal abundance within the bulge. The expansion of the bulge then brings the global fraction of the first strain,~$\bar{f}$, towards~$1/2$. The change in~$\bar{f}$ ceases only when the out-flow boundaries stop moving, which occurs when they are locked between the two neighboring bulges~(see Fig.~\ref{fig:selection}). At this point, the entire front consists of bulges, so~$\bar{f}=1/2$.

\begin{figure*}
\begin{center}
\includegraphics[width=\linewidth]{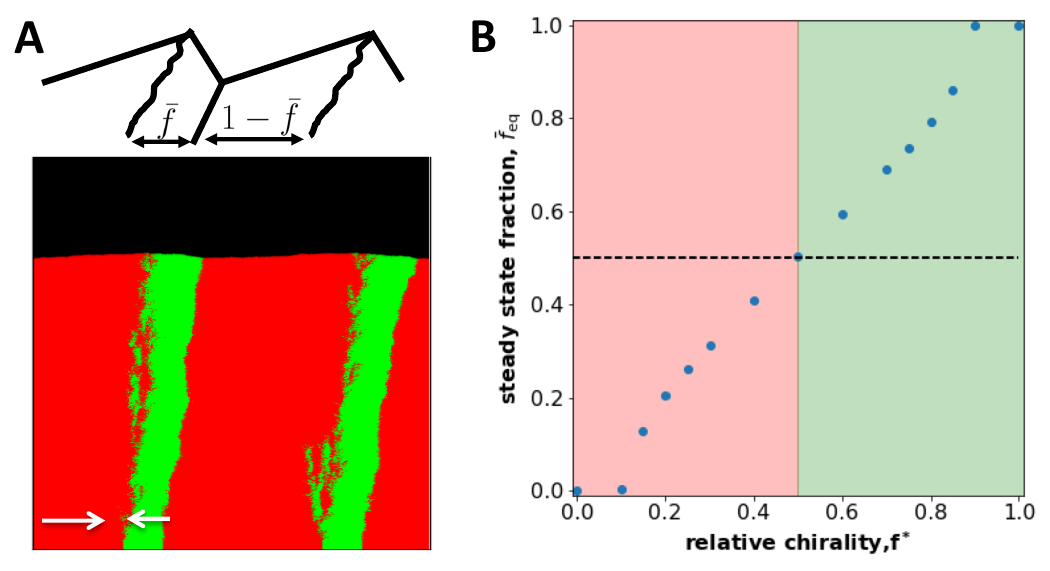}
\caption{\textbf{Equilibrium fractions change with relative chirality.} \textbf{(A)}~shows the steady-state spatial structure for two strains with opposite handedness, but unequal magnitudes of chirality~($f^*\ne1/2$). \textbf{(B)}~shows the relative abundance of the strains as a function of their relative chiralities. This relationship is approximately linear in agreement with Eq.~(\protect{\ref{relative_fraction}}); see also Figs.~S1 and S2. Here,~$m_0=m_s=m_b=m_d=0$,~$g=0.1$ for both strains. The chirality of the strains is shown with thick white arrows. In (A) $m_l^{(2)}=0.0$,~$m_r^{(2)}=0.05$,~$m_l^{(2)}=0.00725$,~$m_r^{(2)}=0.0425$, ~$N=100$. In (B), the chirality of the strains was varied with the difference in their chiralities~$(m_l^{(1)}-m_r^{(1)})- (m_l^{(2)}-m_r^{(2)})=0.1$ and~$m_l+m_r=0.1$ at~$N=400$. Length of lattice was varied to ensure steady state was reached.}
\label{fig:relative_chirality}
\end{center} 
\end{figure*}

The argument above explains the mutual invasion and coexistence for strains with exactly opposite chiralities shown in Fig.~\ref{fig:coexistence}. For~$f^*\ne1/2$, the dynamics are essentially the same with only two minor modifications~(see SI). First, in-flow and out-flow boundaries do not remain stationary within the regions of flat front. Instead, the boundaries move with velocity~$v_{\parallel}=\beta(1/2-f^*)$, which reflects the unequal chiral biases of the two strains. Second, while the bulges remain triangular, they are no longer symmetric relative to the~$y$-axis. The steeper slope occurs on the side that leads the forward motion of the bulge~(Fig.~\ref{fig:relative_chirality}A). 

As before, natural selection occurs due to the expansion of bulges, and the steady state is reached when bulges occupy the entire front. In this state, the relative abundance of the strains is determined by the ratio of bulge slopes, and we find that~(see SI)
\begin{equation}
\label{relative_fraction}
\bar{f}_{\mathrm{eq}} = \left\{
		\begin{aligned}
		& 0,\;\; f^* \le \frac{1}{2} - \frac{\alpha v_0}{2D_h\beta},\\
		& \frac{1}{2} + \frac{\beta D_{h}}{\alpha v_0}\left(f^*- \frac{1}{2}\right),\;\; \left|f^* - \frac{1}{2}\right| < \frac{\alpha v_0}{2D_h\beta} \\
		& 1,\;\; f^* \ge \frac{1}{2} + \frac{\alpha v_0}{2D_h\beta}.\\
		\end{aligned} 
		\right.
\end{equation}

\noindent Here, the middle line describes the relative abundance of the strains when they coexist. The first and the last line describe to the exclusion of the less chiral strain. Exclusion occurs when one of bulge slopes becomes horizontal, and, therefore, no steady state can be reached.

These theoretical conclusions are supported by the simulation results summarized in Fig.~\ref{fig:relative_chirality}B. The data shows a clear transition from coexistence to competitive exclusion and a nearly linear dependence of~$\bar{f}$ on~$f^*$ in the coexistence region as predicted by Eq.~(\ref{relative_fraction}). Quantitative comparison between the theory and the simulations is described in Fig.~S1. Close to the extinction transitions, however, there are noticeable deviations from linearity. Such nonlinearities are typical for non-equilibrium phase transitions and are described by critical exponents~\cite{hinrichsen:review, korolev:mutualism}. To obtain the critical exponent and characterize the nature of the phase transition, one would need to account for the stochastic creation and annihilation of domain boundaries, which we have neglected in our analysis.

The size of the coexistence regions depends on the model parameters~(see Fig.~S2 for a narrower coexistence range). In all of our simulations, we found that the transition between exclusion and coexistence occurs for~$f^*\in(0,1)$. Therefore, the competition between a chiral and a non-chiral strain falls outside the coexistence region, which explains the competitive exclusion shown in Fig.~\ref{fig:chiral_wins}.

\begin{figure*}
\begin{center}
\includegraphics[width=\linewidth]{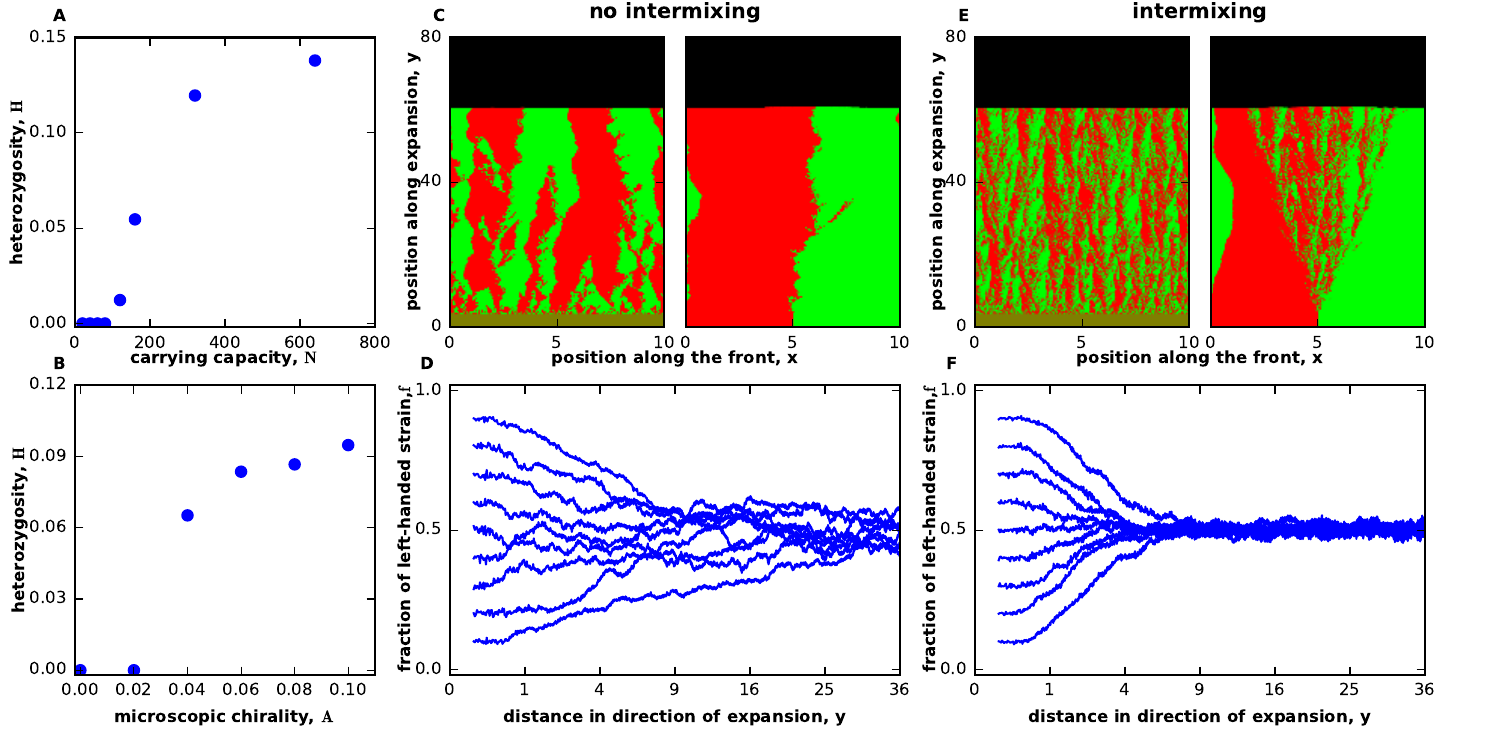}
\caption{\textbf{Transition from segregation to strain intermixing.} The intermixing of the strains was quantified by heterozygosity~$H=\langle 2f(1-f)\rangle$, which is nonzero only when both strains are present at the same spatial location. \textbf{(A,B)}~show that there is a phase transition between an intermixed regime, where~$H$ has a nonzero value at steady state, and a regime, where the strains spatially segregate with ~$H$ vanishing in the long-time limit. The transition is controlled by the relative magnitude of strain chirality and the strength of genetic drift. The latter depends on the number of the organisms at the growing edge. \textbf{(C)} shows the spatial patterns in the demixed regime starting either from a single boundary or from well-mixed initial conditions. \textbf{(D)} shows how the relative abundance of the strains change starting from well-mixed initial conditions and different initial fractions. Note that, even in the demixed regime, there is negative frequency-dependent selection towards coexistence. \textbf{(E,F)} are the same as~(C,D) but in the intermixed regime. All data in this figure are for strains with opposite handedness, but equal magnitude of chirality. Similar results were obtained for unequal magnitudes of chirality; see Fig. S4. Here,~$m_0=m_s=m_b=m_d=0$,~$g=0.1$ for both strains. In panels (A), (C), (D), (E), (F),~$m_l^{(2)}=0.075$,~$m_r^{(2)}=0.025$,~$m_l^{(2)}=0.025$,~$m_r^{(2)}=0.075$ on a lattice of~$1000\textsf{x}8000$ sites with~$N=40$ and ~$N=320$ for no intermixing and intermixing respectively. In (B)~$m_r^{(1)}+m_l^{(1)}=m_r^{(2)}+m_l^{(2)}=0.1$,~$N=200$ was held fixed as chirality was varied. All distances were measured in~$100$ units where~$\Delta \textsf{x}=\Delta \textsf{y}=1$ unit. The right-handed strain is shown in red, and the left-handed strain is shown in green. Note that the equilibration of~$H$ requires that~$f$ is in steady sate; therefore,~$H$ equilibrates more slowly.}
\label{fig:mixing}
\end{center}  
\end{figure*}

\subsection*{Transition to strain intermixing}
So far, our analysis has relied on the existence of sharp boundaries between the strains. Such boundaries appear readily due to genetic drift both in microbial colonies~\cite{hallatschek:sectors} and in our simulations~(Figs.~\ref{fig:chiral_wins},~\ref{fig:bulges_boundaries}-\ref{fig:relative_chirality}). Previously, it has been shown that any non-zero genetic drift prevents diffusive broadening and ensures a finite size of a boundary between two neutral, non-chiral strains~\cite{hallatschek:noisy_fisher}. We found that the behavior is unchanged when the two strains have non-zero, but identical chirality~(Fig.~S3). This observation is not surprising because the front of the colony remains flat, and the chiral motion of the strains can be removed by changing into a reference frame moving along the front. The boundaries also have a finite width~(at least on the time scale of our simulations) when one of the strains is outcompeted due to the differences in growth rates or chirality; see Fig.~S4 and Refs.~\cite{kolmogorov:wave, korolev:sectors}. In this case, the boundary width is controlled both by genetic drift and selection. 

When there is selection towards coexistence, there are two distinct possibilities: Either the boundaries are sharp as in Fig.~\ref{fig:selection} or the boundaries widen over time leading to a intermixed state as in Fig.~\ref{fig:coexistence}. We found that there is a well-defined transition between these two regimes, which is controlled by the strength of chirality and genetic drift~(Figs.~\ref{fig:mixing} and S4). For strong genetic drift or weak chirality, the boundaries between the strains remain sharp, and population dynamics are completely described by the theory developed above. For weak genetic drift or strong chirality, the strains become intermixed. Our main conclusions remain the same even in this regime (see Fig.~S5). In particular, we still observe either coexistence or exclusion depending on the relative chiralities of the strains. The spatial patterns are also similar. A large, non-triangular bulge forms around the intermixed region between the two strains, and small bulges are visible around individual in-flow boundaries~(Figs.~\ref{fig:coexistence} and~\ref{fig:mixing}). 

The transition between demixed and intermixed regimes appears to be continuous~(second order) as seen from Fig.~\ref{fig:mixing}A. Identification of the universality class of this phase transition and the quantitative description of the mixing regime, however, require a careful analysis of the interplay between stochastic and deterministic terms in Eq.~(\ref{effective_theory}) and are left for future work. 

Spatial intermixing could be especially important for species that participate in social interactions such the exchange of metabolites. In such situations, chirality could not only stabilize the coexistence of the species, but also ensure that they are sufficiently close to each other. When no special mechanism exists to ensure spatial proximity, mutualistic interactions can be easily destroyed by genetic drift~\cite{korolev:mutualism, momeni:intermixing, muller:mutualism, menon:diffusion}. Hence, some microbes may rely on different or fluctuating handedness to ensure that the separation between the species does not exceed the maximal distance over which they can interact. 


\section*{Discussion}

Selection for a particular chirality may seem impossible because an object and its mirror image have identical physical properties. This apparent paradox is however easily resolved by noticing that natural selection always favors a change in chirality relative to that of the ancestral population rather than an absolute, pre-defined value of chirality.

A vivid example of how evolution drives a change in chirality comes from \textit{Satsuma} snails. Most species in the \textit{Satsuma} genus are dextral~(clockwise coiled), but they often have sister species that are almost identical except for the opposite direction of coiling~\cite{hoso:snails}. These sinister~(counterclockwise coiled) species enjoy a distinct selective advantage because they are essentially resistant to the predation by \textit{Pareatidae iwasakii}, a snake that is common in the rage of \textit{Satsuma}~\cite{hoso:snails}. Resistance to predation comes from the left-right asymmetry in the jaw of \textit{P.~iwasakii}, which has adapted to the coiling direction of its most common prey. Similarly, a reversal of handedness provides protection to \textit{Arthrospira}, cyanobacteria that forms helical trichomes, from the predation by a ciliate~\cite{whitton:helix_reversal, young:shape}. In both examples, the mutants enjoy the advantage of being in the minority. This mechanism does not require the presence of a predator and can occur due to a large number of factors. For example, a mutant with chiral motility may spatially segregate from the rest of the population and thereby escape from an intense competition for resources~\cite{bechinger:active_matter_review}.    

Our main finding is that selection for chirality can also be mediated by the formation of non-trivial spatial patterns. Mismatch in the chiral bias makes cells move towards each other near in-flow boundaries and away from each other near out-flow boundaries. As a result, the colony edge becomes populated with bulges and dips, which grow over time and alter the relative abundance of the strains. One consequence of these dynamics is that it pays off to be different from the majority of the population: A mutant with the opposite handedness can invade when rare and stably coexist with the ancestor due to negative frequency-dependent selection. For strains with the same handedness, the more chiral strain typically wins the competition because it creates a one-sided bulge that overgrows the less chiral strain. Thus, we identified a distinct selection mechanism that can explain both the evolution toward stronger chirality and sudden reversal of handedness. The predicted effects of chirality are observable even in the presence of moderate growth rate differences and can be tested experimentally by comparing the competition between cells with different chirality to our predictions for colony shape and composition.

Selection for chirality could also come from the indirect benefits of the emergent spatial pattern. One possibility is that pointed bulges might facilitate the invasion of host tissue or other environments. The other possibility is that strain intermixing could promote social interactions that rely on cell contact or the exchange of diffusible metabolites. We found that intermixing between the strains with opposite handedness is stable only when their chiralities exceed a certain threshold. Below this threshold, genetic drift creates macroscopic sectors that grow over time and spatially segregate the strains. As a result of this process, social interactions are either suppressed or completely abolished~\cite{korolev:mutualism, momeni:intermixing, muller:mutualism, menon:diffusion}.

All of our results can be explained by a simple effective theory that describes population dynamics in terms of colony shape and composition. This description is simpler and much more intuitive than the full two-dimensional growth encoded by reaction-diffusion equations or other mechanistic models. Therefore, our theory could provide a valuable framework to study competition and cooperation in compact aggregates such as microbial colonies and cancer tumors. 

Shape undulations inevitably occur when aggregates contain strains that grow at different rates~\cite{korolev:sectors, momeni:cheater}. So far, most theoretical studies have neglected this complexity and assumed that colonies have a flat front~\cite{korolev:review, korolev:mutualism, menon:diffusion}. Front undulations, however, are known to profoundly change the nature of competition, for example, by allowing the regions with cooperating strains to overgrow the regions where cooperation has been lost~\cite{momeni:cheater, lavrentovich:mutualism}. Our theory is an important first step towards understanding this interplay between evolution in compact aggregates and their shape. 

The effective theory also provides an interesting extension of the KPZ equation to systems that break the mirror symmetry. Such symmetry breaking could occur in a variety of systems within the KPZ universality class, for example, during the simultaneous deposition of two homophilic molecules with opposite handedness.

In summary, we have identified a new mechanism of selection for chirality and developed a theory to explain it. Our findings describe the chirality of cells while most of the previous work focused on the emergence of homochirality in biological molecules~\cite{kusmin:parity, quack:parity, blackmond:homochirality_review, frank:chirality, goldenfeld:chirality}. Unlike the frozen homochirality of nucleotides and amino acids, the chirality of cells continues to evolve, often on the time scale of a few generations~\cite{whitton:helix_reversal, young:shape, wan:cancer_chiral, ben_jacob:engineering}. Our work suggests that some changes in cellular chirality could be adaptive and, therefore, deserve further study.

\section*{Methods}

Lattice-based simulations were performed on a two-dimensional rectangular grid with periodic boundary conditions. The lattice spacings~$\Delta x$ and~$\Delta y$ were both set to~$1$. The length of each time step~$\Delta t$ was set to~$1$ as well. Each time step, we first performed a deterministic update to account for growth and migration and then performed a stochastic update to account for demographic fluctuations and genetic drift. To speed up simulations, only a small rectangular region at the expanding edge was updated while the bulk remained frozen. This did not not affect the results since both migration and growth are zero in the colony bulk. The heterozygosity reported in Fig.~\ref{fig:mixing}, was calculated within this rectangular region at each time step.

\subsection*{Deterministic update}
During the deterministic update, we computed an auxiliary quantity~$\rho^{(\alpha)}$ equal to the expected value of~$n^{(\alpha)}$ at the next time step:

\begin{equation}
\rho^{(\alpha)}(t,x,y) = n^{(\alpha)}(t,x,y) + G^{(\alpha)}(t,x,y)\Delta t + M^{(\alpha)}(t,x,y)\Delta t.
\end{equation}

The growth term~$G$ describes the increase in the population density due to logistic growth:

\begin{equation}
G^{(\alpha)}(t,x,y)=g^{(\alpha)}n^{(\alpha)}(t,x,y)\left(1-\frac{n(t,x,y)}{N}\right),
\end{equation}

\noindent where~$n=n^{(1)}+n^{(2)}$ is the total population size, and~$N$ is the carrying capacity. The growth rates~$g^{(\alpha)}$ were typically the same for the two strains.

The migration term~$M$ describes the change in the population size due to migration:
\begin{equation}
\begin{aligned}
M^{(\alpha)}(t,x,y)= & -m^{(\alpha)}_{(x,y)\to(x+\Delta x,y)} - m^{(\alpha)}_{(x,y)\to(x,y+\Delta y)} - m^{(\alpha)}_{(x,y)\to(x-\Delta x,y)} - m^{(\alpha)}_{(x,y)\to(x,y-\Delta y)}\\
& + m^{(\alpha)}_{(x+\Delta x,y)\to(x,y)} + m^{(\alpha)}_{(x,y+\Delta y)\to(x,y)} + m^{(\alpha)}_{(x-\Delta x,y)\to(x,y)} + m^{(\alpha)}_{(x,y-\Delta y)\to(x,y)},
\end{aligned}
\end{equation}

\noindent where~$m^{(\alpha)}_{(x_1,y_1)\to(x_2,y_2)}\Delta t$ is the expected number of migrants of strain~$\alpha$ from the site at~$(x_1,y_1)$ into the site at~$(x_2,y_2)$. The first four terms in the equation describe migration out of the lattice site~$(x,y)$ and the last four terms describe the migration into the lattice site~$(x,y)$. Note that the number of cells leaving a particular site into the direction of its nearest neighbor is equal to the number of cells arriving into that neighboring site, i.e. migration conserves the number of cells.

The migration fluxes~$m^{(\alpha)}_{(x_1,y_1)\to(x_2,y_2)}$ were nonzero only between the four nearest neighbors and were defined as follows

\begin{equation}
\label{migration_rates}
\begin{aligned}
& m^{(\alpha)}_{(x,y)\to(x+\Delta x,y)} = n^{(\alpha)}(t,x,y)\left(1-\frac{n(t,x+\Delta x,y)}{N}\right)\times\\&\left( m_0^{(\alpha)} + m_{\mathrm{s}}^{(\alpha)}\frac{n(t,x,y)}{N} + m_{\mathrm{d}}^{(\alpha)}\frac{n(t,x+\Delta x,y)}{N} + m_{\mathrm{l}}^{(\alpha)}\frac{n(t,x,y+\Delta y)}{N} + m_{\mathrm{b}}^{(\alpha)}\frac{n(t,x-\Delta x,y)}{N} + m_{\mathrm{r}}^{(\alpha)}\frac{n(t,x,y-\Delta y)}{N} \right),\\
& m^{(\alpha)}_{(x,y)\to(x,y+\Delta y)} = n^{(\alpha)}(t,x,y)\left(1-\frac{n(t,x,y+\Delta y)}{N}\right)\times\\&\left( m_0^{(\alpha)} + m_{\mathrm{s}}^{(\alpha)}\frac{n(t,x,y)}{N} + m_{\mathrm{d}}^{(\alpha)}\frac{n(t,x,y+\Delta y)}{N} + m_{\mathrm{l}}^{(\alpha)}\frac{n(t,x-\Delta x,y)}{N} + m_{\mathrm{b}}^{(\alpha)}\frac{n(t,x,y-\Delta y)}{N} + m_{\mathrm{r}}^{(\alpha)}\frac{n(t,x+\Delta x,y)}{N} \right),\\
& m^{(\alpha)}_{(x,y)\to(x-\Delta x,y)} = n^{(\alpha)}(t,x,y)\left(1-\frac{n(t,x-\Delta x,y)}{N}\right)\times\\&\left( m_0^{(\alpha)} + m_{\mathrm{s}}^{(\alpha)}\frac{n(t,x,y)}{N} + m_{\mathrm{d}}^{(\alpha)}\frac{n(t,x-\Delta x,y)}{N} + m_{\mathrm{l}}^{(\alpha)}\frac{n(t,x,y-\Delta y)}{N} + m_{\mathrm{b}}^{(\alpha)}\frac{n(t,x+\Delta x,y)}{N} + m_{\mathrm{r}}^{(\alpha)}\frac{n(t,x,y+\Delta y)}{N} \right),\\
& m^{(\alpha)}_{(x,y)\to(x,y-\Delta y)} = n^{(\alpha)}(t,x,y)\left(1-\frac{n(t,x,y-\Delta y)}{N}\right)\times\\&\left( m_0^{(\alpha)} + m_{\mathrm{s}}^{(\alpha)}\frac{n(t,x,y)}{N} + m_{\mathrm{d}}^{(\alpha)}\frac{n(t,x,y-\Delta y)}{N} + m_{\mathrm{l}}^{(\alpha)}\frac{n(t,x+\Delta x,y)}{N} + m_{\mathrm{b}}^{(\alpha)}\frac{n(t,x,y+\Delta y)}{N} + m_{\mathrm{r}}^{(\alpha)}\frac{n(t,x-\Delta x,y)}{N} \right),\\
\end{aligned}
\end{equation}

\noindent where the factors of~$n^{(\alpha)}$ ensure that the number of migrants is proportional to the local abundance of the strain, and the factors of~$1-\frac{n}{N}$ ensure that migration cannot occur into occupied lattice sites. As a result of these choices, the spatial distribution of the strains remains ``frozen'' behind the growing front just as in microbial colonies, where the growth in the bulk of the colony is suppressed. The last factor in each of the equations describes the dependence of the migration rates on the local population population density and its spatial gradients; this can be seen by expanding population densities into Taylor series. 

\begin{figure*}
\begin{center}
\includegraphics[width=\linewidth]{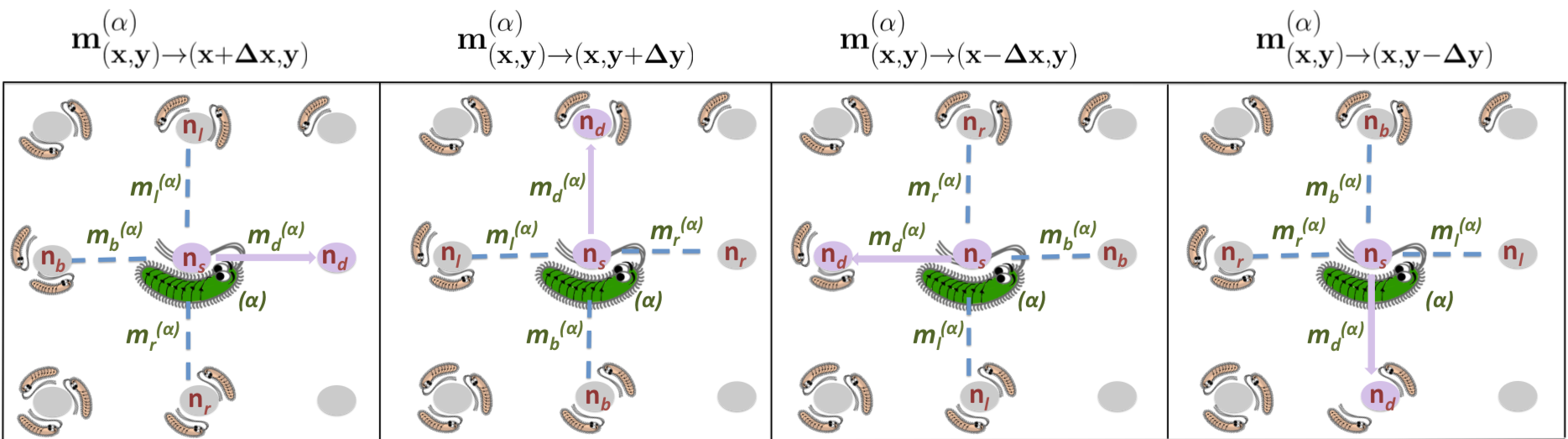}
\caption{\textbf{Isotropic, but chiral migration in a lattice-based model.} The panels illustrate the computation of migration fluxes~$m^{(\alpha)}_{(x_1,y_1)\to(x_2,y_2)}$ given by Eq.~(\ref{migration_rates}) for different orientations of the migration direction relative to the lattice. Note that the choice of the migration coefficients is always made relative to the direction of migration and not relative to the coordinate system.}
\label{fig:lattice_model}
\end{center}  
\end{figure*}

Note that our definitions preserve the equivalence of all four lattice direction because the migration coefficients are chosen according to the position of the lattice sites relative to the direction of the migration rather than relative to a particular lattice direction; see Fig.~\ref{fig:lattice_model}. To emphasize this fact, we use the index labels that refer to source site, destination site, left site, back site, and right site---all specified with respect to the migration direction. For simplicity, we limited the dependence on~$n$ to the lowest order of the Taylor expansion that is sufficient to produce chiral growth. 

The relationship between the model parameters and the coefficients in the continuum description is provided below:
\begin{equation}
\begin{aligned}
	& D^{(\alpha)}(n) = \left[ m_0 ^{(\alpha)} + \frac{n}{N} \left(m_{\mathrm{s}}^{(\alpha)}  +m_{\mathrm{d}}^{(\alpha)}  +m_{\mathrm{l}}^{(\alpha)}  +m_{\mathrm{b}}^{(\alpha)}  +m_{\mathrm{r}} ^{(\alpha)}    \right)        \right]   \left(1-\frac{n}{N} \right)\frac{\Delta x^2}{\Delta t}\\
	& S^{(\alpha)}(n)= \left[2\left(m_{\mathrm{b}}^{(\alpha)}  -m_{\mathrm{d}}^{(\alpha)}\right)\left(1-\frac{n}{N}\right) +\left(m_0^{(\alpha)}+m_{\mathrm{s}}^{(\alpha)}  +m_{\mathrm{d}}^{(\alpha)}  +m_{\mathrm{l}}^{(\alpha)}  +m_{\mathrm{b}}^{(\alpha)}  +m_{\mathrm{r}} ^{(\alpha)}    \right)\right]\frac{\Delta x^2}{\Delta t}\\
	& A^{(\alpha)}(n)= 2\left(m_{\mathrm{l}}^{(\alpha)}  -m_{\mathrm{r}}^{(\alpha)}\right)\left(1-\frac{n}{N}\right)\frac{\Delta x^2}{\Delta t}\\
\label{eq:parameter_mapping}
\end{aligned}
\end{equation}

From Eq.~(\ref{eq:parameter_mapping}), it is clear that~$A^{(\alpha)}$ depends on~$m_{\mathrm{l}}^{(\alpha)}-m_{\mathrm{r}}^{(\alpha)}$ while~$D^{(\alpha)}$ depends on~$m_{\mathrm{l}}^{(\alpha)}+m_{\mathrm{r}}^{(\alpha)}$. Thus, one can vary the chirality of a strain without affecting its motility. We used this freedom to isolate the effects of chirality from other components of strain fitness in most of our simulations by keeping~$m_{\mathrm{l}}^{(\alpha)}+m_{\mathrm{r}}^{(\alpha)}$ fixed.

\subsection*{Stochastic update}
The stochastic update consisted of two rounds of binomial sampling.

The first round accounted for the demographic fluctuations in the total population size. We drew~$n(t+\Delta t,x,y)$ from a binomial distribution with~$N$ trials and~$(\rho^{(1)}(t,x,y)+\rho^{(2)}(t,x,y))/N$ probability of success. This procedure ensures that (i)~the expectation value of~$n$ is consistent with the deterministic dynamics, (ii)~the size of a typical fluctuation scales as~$\sqrt{n}$ for~$n\ll N$, and (iii)~the population size never exceeds the carrying capacity~$N$. 

The second round accounted for genetic drift. We drew~$n^{(1)}(t+\Delta t,x,y)$ from a binomial distribution with~$n(t+\Delta t,x,y)$ trials and~$\rho^{(1)}(t,x,y)/(\rho^{(1)}(t,x,y)+\rho^{(2)}(t,x,y))$ probability of success. The abundance of the other strain was set to~$n^{(2)}(t+\Delta t,x,y)=n(t+\Delta t,x,y) - n^{(1)}(t+\Delta t,x,y)$. This stochastic update does not change the relative fractions of the two strains on average, and the typical fluctuation in the relative abundance of the strains scales as~$\sqrt{n}$.

\subsection*{Off-lattice simulations}
To ensure that our results do not arise because of the lattice effects, we developed off-lattice simulations of our reaction-diffusion model. In these simulations, cells reproduced stochastically depending on the local population density and performed short-range jumps. The magnitude of the jump was controlled by the population density and the direction of the jump depended on the local density gradient and the chirality of the cell. The functional forms of the growth rates and the jump kernels are provided in the SI. 

Off-lattice simulations confirmed the predictions of our theory and lattice-based simulations. Specifically, we observed stabilizing selection and the formation of bulges between strains with opposite handedness. These results are shown in the SI. Because of computational efficiency, most of the analysis was carried out using lattice-based simulations.

\section*{Acknowledgments}
Simulations were carried out on the Boston University Shared Computing Cluster. This work was partially supported by the Cottrell Scholar Award by the Research Corporation for Science Advancement, a grant from the Simons Foundation, and by a grant from Moore foundation.

\newpage
\section*{Supporting information}

\renewcommand{\theequation}{S\arabic{equation}}
\renewcommand{\thefigure}{S\arabic{figure}}
\renewcommand{\thesection}{S\arabic{section}}
\renewcommand{\thetable}{S\arabic{table}}
\setcounter{equation}{0}
\setcounter{figure}{0}
\setcounter{section}{0}
\setcounter{table}{0}


\section{Formulation of the reaction-diffusion model}
The goal of this section is to formulate a reaction-diffusion model that describes competition between two chiral strains in growing colonies. Quite generally, a reaction-diffusion model can be stated as follows

\begin{equation}
\label{movement_growth}
\frac{\partial n^{(\alpha)}}{\partial t} = g^{(\alpha)}n^{(\alpha)} - \boldsymbol{\nabla}\boldsymbol{\cdot}\boldsymbol{J}^{(\alpha)},
\end{equation}

\noindent where~$n^{(\alpha)}(t,x,y)$ is the population density of the strain of type~$\alpha$ at time~$t$ and position~$(x,y)$;~$g^{(\alpha)}$ is the growth rate;~and $J^{(\alpha)}$ is the population flux due to movement. Both~$g^{(\alpha)}$ and~$J^{(\alpha)}$ could depend on the population densities of all strains, but not on~$t$,~$x$, or~$y$ because we assume spatial and temporal homogeneity. Here and below, bold symbols denote two dimensional vectors and $\boldsymbol{\nabla}$ denotes the gradient operator. 

The chirality of the strains manifests through their active or passive movement and is encoded in~$\boldsymbol{J}^{(\alpha)}$. Assuming that~$n^{(\alpha)}$ are the only relevant dynamical variables, one has only two vectors that can determine the orientation of~$\boldsymbol{J}^{(\alpha)}$. These are~$\boldsymbol{\nabla}_{i}n^{(1)}$ and~$\boldsymbol{\nabla}_{i}n^{(2)}$ or equivalently the gradient of the total population density~$n=n^{(1)}+n^{(2)}$ and the gradient of the population density of a specific strain. Thus, to the first order in the gradient expansion one can write the population fluxes as

\begin{equation}
\boldsymbol{J}^{(\alpha)}_{i} = -D^{(\alpha)} \boldsymbol{\nabla}_{i}n^{(\alpha)} - n^{(\alpha)}\sum_{j}\left(S^{(\alpha)}\delta_{ij}- A^{(\alpha)} \epsilon_{ij}\right)\boldsymbol{\nabla}_{j}n ,
\label{flux_decomposition}
\end{equation}

\noindent where the indexes denote the Cartesian components of vectors;~$\delta_{ij}$ is the unit tensor; and~$\epsilon_{ij}$ is the totally antisymmetric tensor, also known as Levi-Civita symbol. The first term in the equation accounts for diffusion-like movement. The second term describes the advection of the strain density by local motion within the aggregate. This motion arises due to gradients of mechanical pressure, chemotaxis, and other factors. The direction of the advection velocity is determined by the gradient of the total population density. The term proportional to~$S^{(\alpha)}$ describes the flux along the gradient, which is the only possibility for non-chiral organisms. The term proportional to~$A^{(\alpha)}$ describes the motion perpendicular to~$\boldsymbol{\nabla}n$ due to chirality. The above decomposition of the drift term into symmetric and antisymmetric parts is always possible because~$\delta_{ij}$ and~$\epsilon_{ij}$ are the only rotationally invariant tensors in two dimensions. Note that we do not include a separate term proportional to~$\epsilon_{ij}\boldsymbol{\nabla}_{j}n^{(\alpha)}$ because its contribution to Eq.~(\ref{movement_growth}) is indistinguishable from the contribution of the term proportional to~$A^{(\alpha)}$. In the following, we refer to~$A^{(\alpha)}$ as the microscopic ``chiralities'' of the strains. The sign of these chiralities describes the overall chirality of the strain: positive for strains that tend to move to the left relative to the expansion direction and negative for strains that tend to move to the right relative to the expansion direction. The magnitudes of the chiralities reflect the strength of the movement biases. 

Since our primary goal is to understand the effects chirality, we exclude from consideration all differences between the strains that are possible even when the strains have the same chirality. In particular, we assume that~$g^{(\alpha)}$,~$D^{(\alpha)}$,~$S^{(\alpha)}$ are equal for the two strains and depend only on the total population density. The coefficients describing chirality could be different between the two strains, but we still assume that~$A^{(\alpha)}$ depend only on~$n$. Note that local interaction between the chiral strains, e.g. via alignment, could results in the dependence of~$D^{(\alpha)}$,~$S^{(\alpha)}$, and~$A^{(\alpha)}$ on the relative fractions of the species. It is easy to check, however, that this dependence generates terms in Eq.~(\ref{movement_growth}) that are invariant under mirror symmetry. When such terms arise due to chirality, they must scale at least quadratically in the difference between the chiralities of the two strains because this difference changes sign under the application of the mirror-symmetry transformation.\footnote{We discuss this point further in the section devoted to the phenomenological derivation of the effective equations for the shape and composition of the edge of a growing colony.} Further, the effects of mirror-symmetric terms are less interesting because they could arise even in the absence of chirality.  

After the simplifications just described, the equation governing population dynamics reads

\begin{equation}
\label{strain_dynamics}
\frac{\partial n^{(\alpha)}}{\partial t} = g(n)n^{(\alpha)} + \boldsymbol{\nabla}\left(D(n)\boldsymbol{\nabla}n^{(\alpha)}\right) + \boldsymbol{\nabla}\left(n^{(\alpha)}S(n)\boldsymbol{\nabla}n\right)  - A^{(\alpha)}(n)\sum_{ij} \epsilon_{ij} (\boldsymbol{\nabla}_i n^{(\alpha)}) (\boldsymbol{\nabla}_{j}n).
\end{equation}

\noindent Note that~$A^{(\alpha)}(n)$ can be taken outside of the derivatives because~$\sum_{ij} \epsilon_{ij} \boldsymbol{\nabla}_i n\boldsymbol{\nabla}_{j}n=0$.

Althoug the analysis can be carried out for arbitrary~$g(n)$,~$D(n)$,~$S(n)$, and~$A^{(\alpha)}(n)$. We will make a few assumptions that significantly simplify the discussion and make it easier to connect the theory to experimental results shown in Fig. 1 in the main text. First, we assume that as the population grows it reaches a stable state at~$n=N$, which correspond to the carrying capacity of the population. Mathematically, this assumption can be stated as~$g(N)=0$ and~$\left.\frac{dg}{dn}\right|_{n=N}<0$. Second, we assume that all migration ceases once the population reaches the carrying capacity. That is~$D(N)$,~$S(N)$, and~$A^{(\alpha)}(N)$ are zero. This assumption reflects the fact that no dynamics occur behind the expansion front in the experiments shown in Fig.1 in the main text. When this assumption does not hold, the results derived below apply just at the front of the population and do not describe the dynamics in the bulk of the colony. The third assumption is that the population expands as a ``pushed'' reaction-diffusion wave \textit{sensu} Ref.~\cite{saarloos:review}. This means that the expansion velocity is greater than the expansion velocity obtained by linearizing Eq.~(\ref{n_dynamics}) for small~$n$, which is known as the Fisher or linear spreading velocity~\cite{saarloos:review}. Pushed waves occurs when at least one function out of~$D(n)$,~$S(n)$, and~$g(n)$ exhibits a substantial increase with~$n$ relative to its value at~$n=0$. Because these functions vanish at~$n=N$, at least one of the functions must be non-monotonic in~$n$.

The main reason why we assume that the expansions are pushed is that Eq.~(\ref{strain_dynamics}) predicts no chiral phenomena for pulled waves. Indeed, the dynamics of pulled waves are described by the equation linearized around~$n=0$, in which the chiral term vanishes because it is quadratic in~$\boldsymbol{\nabla}n$. 

To understand the competition between the strains, it is convenient to recast Eq.~(\ref{strain_dynamics}) in terms of the total population density and the relative fraction of the first strain~$f=n^{(1)}/n$. This change of variables results in the following equations

\begin{equation}
\label{n_dynamics}
\frac{\partial n}{\partial t} = g(n)n + \boldsymbol{\nabla}\left[(D(n)+nS(n))\boldsymbol{\nabla}n\right]  -  \left[A^{(1)}(n) - A^{(2)}(n) \right] n \sum_{ij} \epsilon_{ij} (\boldsymbol{\nabla}_i f) (\boldsymbol{\nabla}_{j}n),
\end{equation}

and

\begin{equation}
\label{f_dynamics}
\begin{aligned}
\frac{\partial f}{\partial t} = & \boldsymbol{\nabla}\left[D(n)\boldsymbol{\nabla}f\right] + \left[ \frac{2D(n)}{n} + S(n)\right](\boldsymbol{\nabla} n)(\boldsymbol{\nabla}f) \\  &- \left[A^{(1)}(n) - A^{(2)}(n) \right] \left[\frac{A^{(1)}(n)}{A^{(1)}(n) - A^{(2)}(n)} - f\right] \sum_{ij} \epsilon_{ij} (\boldsymbol{\nabla}_i f) (\boldsymbol{\nabla}_{j}n).
\end{aligned}
\end{equation}

\section{Derivation of the effective theory of front shape and composition}

In this section, we replace the detailed model of growth in two spatial dimensions by an approximate model in terms of the shape and composition of the population front. The motivation for this approximation is clear: Away from the edge, the population densities are either zero or stationary because $n$~is at the carrying capacity. The approach presented below is a type of dimensional reduction that eliminates one of the spatial variables and is reminiscent of the moving-boundary approximation developed in other contexts~\cite{saarloos:review}. The final result is an effective description in terms of the position of the population front and the composition of the population at the front.

For simplicity, we consider the colony front that is mostly parallel to the $x$-axis so that the shape of the front can be described by its~$y$ coordinate as a function of~$x$. We denote this dependence as~$h(t,x)$ and refer to front position as height in analogy with surface growth phenomena. The composition of the population at the growing edge is described by the fraction of the first strain~$f(t,x)$. Below, we derive the equations governing the dynamics of~$h(t,x)$ and~$f(t,x)$ starting from Eqs.~(\ref{n_dynamics}) and~(\ref{f_dynamics}). The main two assumptions underlying this derivation is that the front undulations are small and that there is little variation in~$f$ across the thickness of the colony edge. These assumptions require that the expansion is pushed and that the difference in the chiralities of the two strains is small.

\subsection*{Equation for front shape,~$\boldsymbol{h(t,x)}$}
The equation for~$h(t,x)$ follows from Eq.~(\ref{n_dynamics}), which we re-write as

\begin{equation}
\label{n_dynamics_compact}
\frac{\partial n}{\partial t} = g(n)n + \boldsymbol{\nabla}\left(D_e\boldsymbol{\nabla}n\right)  - \mathcal{A} \sum_{ij} \epsilon_{ij} (\boldsymbol{\nabla}_i f) (\boldsymbol{\nabla}_{j}n)
\end{equation}

\noindent by introducing a more compact notation:~$D_{e} = D + nS$ and~$\mathcal{A} =  (A^{(1)} - A^{(2)}) n $.

To carry out the dimensional reduction described above, we assume that~$n(t,x,y)=n(\zeta)$, where~$\zeta$ is the distance between~$(x,y)$ and the line defined by the new dynamical variable~$h(t,x)$. The new variable~$h(t,x)$ represents the $y$~coordinate of the population front for a given~$x$ and could, for example, be defined such that~$n(t,x,h(t,x))=1/2\max_{(x,y)}\{n(t,x,y)\}$. For a flat front parallel to the $x$-axis,~$\zeta=y-h(t,x)$, but for a tilted front~$\zeta$ is smaller and is given by the projection of~$y-h$ on the direction perpendicular to the front. In the following, we assume that~$\frac{\partial h}{\partial x}\ll 1$ and keep only the terms up to the second order in~$\frac{\partial }{\partial x}$. Under these assumptions, the tilt of the front is given by~$\frac{\partial h}{\partial x}$ and the expression for~$\zeta$ reads

\begin{equation}
\label{zeta}
\zeta = [y-h(t,x)]\left[1-\frac{1}{2}\left(\frac{\partial h}{\partial x}\right)^2\right].
\end{equation}

We now evaluate all the terms that enter Eq.~(\ref{n_dynamics_compact}) up to the second order in~$\frac{\partial }{\partial x}$:

\begin{equation}
\begin{aligned}
& \frac{\partial n}{\partial t} = - \frac{dn}{d\zeta}\frac{\partial h}{\partial t}\left[1-\frac{1}{2}\left(\frac{\partial h}{\partial x}\right)^2\right],\\
& \boldsymbol{\nabla}n = \frac{dn}{d\zeta}\left[1-\frac{1}{2}\left(\frac{\partial h}{\partial x}\right)^2\right]\left(-\frac{\partial h}{\partial x}, 1\right),\\
& (\boldsymbol{\nabla}n)^2 = \left(\frac{dn}{d\zeta}\right)^2,\\
& \boldsymbol{\nabla}^2 n = \frac{d^2n}{d\zeta^2} - \frac{dn}{d\zeta}\frac{\partial^2 h}{\partial x^2},\\
& \sum_{ij} \epsilon_{ij} (\boldsymbol{\nabla}_i f) (\boldsymbol{\nabla}_{j}n) = \frac{\partial f}{\partial x}\frac{dn}{d\zeta},\\
\end{aligned}
\label{h_calculation_terms}
\end{equation}

\noindent where, in the last expression, we assumed that~$f$ remains constant in the direction perpendicular to the front. Upon substituting Eq.~(\ref{h_calculation_terms}) into Eq.~(\ref{n_dynamics_compact}), we obtain

\begin{equation}
\label{reduced_ode}
\frac{d}{d\zeta}\left(D_e\frac{dn}{d\zeta}\right) + \frac{dn}{d\zeta}\frac{\partial h}{\partial t}\left[1-\frac{1}{2}\left(\frac{\partial h}{\partial x}\right)^2\right] + gn   - \frac{dn}{d\zeta}\left(D_e\frac{\partial^2 h}{\partial x^2} + \mathcal{A}\frac{\partial f}{\partial x}\right) = 0.
\end{equation}

To understand this equation, let us first consider a flat front parallel to the $x$-axis~($\frac{\partial h}{\partial x} = 0$) with no spatial variation in~$f$. Such a front moves along the $y$-axis with a constant velocity~$v_0$, i.e.~$\frac{\partial h}{\partial t} = v_0$. Hence, Eq.~(\ref{reduced_ode}) reduces to 

\begin{equation}
\label{reduced_ode_wave}
\frac{d}{d\zeta}\left(D_e\frac{dn}{d\zeta}\right) + v_0 \frac{dn}{d\zeta} + gn  = 0,
\end{equation}

\noindent which is the standard equation for the population density profile of a reaction-diffusion wave. The solution of this equation has to satisfy three constraints:~$\lim_{\zeta\to-\infty}n=0$,~$\lim_{\zeta\to+\infty}n=N$, and~$n(0)=N/2$, which eliminates the translational degree of freedom by specifying the position of a particular population density. Since Eq.~(\ref{reduced_ode_wave}) is second order, the three constraints cannot be satisfied simultaneously for an arbitrary~$v_0$. Thus, Eq.~(\ref{reduced_ode_wave}) determines both the profile shape~$n(\zeta)$ and the expansion velocity~$v_0$. We note that this simple argument applies only to pushed waves, and additional analysis is needed to determine the expansion velocity of pulled waves~\cite{saarloos:review, murray:mathematical_biology, birzu:semi_pushed}.

We now return to the full Eq.~(\ref{reduced_ode}) and analyze it in the limit of~$\frac{\partial h}{\partial t}$,~$\frac{\partial h}{\partial x}$,~$\frac{\partial^2 h}{\partial x^2}$, and~$\frac{\partial f}{\partial x}$ being approximately constant. The first three terms in Eq.~(\ref{reduced_ode}) are then equivalent to Eq.~(\ref{reduced_ode_wave}) while the remaining terms constitute a small perturbation that depends on~$n$. This perturbation leads to a correction to the expansion velocity, which can be evaluated using the results from Refs.~\cite{gottwald:melnikov, goldenfeld:structural_stability, birzu:semi_pushed, matin:cooperative_migration}. The expression for the expansion velocity can then be equated to the coefficient in front of~$\frac{dn}{d\zeta}$ in the third term of Eq.~(\ref{reduced_ode}), which leads to

\begin{equation}
\label{v_step}
\frac{\partial h}{\partial t}\left(1-\frac{1}{2}\left(\frac{\partial h}{\partial x}\right)^2\right) = v_0 + \frac{\int_{-\infty}^{+\infty}\left[D_e(n_0(\zeta))\frac{\partial^2 h}{\partial x^2} + \mathcal{A}(n_0(\zeta))\frac{\partial f}{\partial x}\right]D_e(n_0(\zeta)) \left(\frac{dn_0(\zeta)}{d\zeta}\right)^2e^{v_0\int_{0}^{\zeta}\frac{d\tilde{\zeta}}{D_e(n_0(\tilde{\zeta}))}}d\zeta}{\int_{-\infty}^{+\infty}D_e(n_0(\zeta)) \left(\frac{dn_0(\zeta)}{d\zeta}\right)^2e^{v_0\int_{0}^{\zeta}\frac{d\tilde{\zeta}}{D_e(n_0(\tilde{\zeta}))}}d\zeta},
\end{equation}

\noindent where~$n_0(\zeta)$ is the density profile satisfying Eq.~(\ref{reduced_ode_wave}). Since~$\frac{\partial^2 h}{\partial x^2}$ and~$\frac{\partial f}{\partial x}$ do not depend on~$\zeta$, they can be taken outside of the integrals and we obtain the following equation for~$h(t,x)$:

\begin{equation}
\label{h_equation}
\frac{\partial h}{\partial t} = v_0 + \frac{v_0}{2}\left(\frac{\partial h}{\partial x}\right)^2 + D_h \frac{\partial^2 h}{\partial x^2} + \alpha\frac{\partial f}{\partial x},
\end{equation}

\noindent where

\begin{equation}
\label{D_h}
D_h = \frac{\int_{-\infty}^{+\infty}D^2_e(n_0(\zeta))\left(\frac{dn_0(\zeta)}{d\zeta}\right)^2e^{v_0\int_{0}^{\zeta}\frac{d\tilde{\zeta}}{D_e(n_0(\tilde{\zeta}))}}d\zeta}{\int_{-\infty}^{+\infty}D_e(n_0(\zeta)) \left(\frac{dn_0(\zeta)}{d\zeta}\right)^2e^{v_0\int_{0}^{\zeta}\frac{d\tilde{\zeta}}{D_e(n_0(\tilde{\zeta}))}}d\zeta},
\end{equation}

\noindent and

\begin{equation}
\label{alpha}
\alpha = \frac{\int_{-\infty}^{+\infty}\left\{  [A^{(1)}(n_0(\zeta))- A^{(2)}(n_0(\zeta))] n_0(\zeta) \right\} D_e(n_0(\zeta)) \left(\frac{dn_0(\zeta)}{d\zeta}\right)^2e^{v_0\int_{0}^{\zeta}\frac{d\tilde{\zeta}}{D_e(n_0(\tilde{\zeta}))}}d\zeta}{\int_{-\infty}^{+\infty}D_e(n_0(\zeta)) \left(\frac{dn_0(\zeta)}{d\zeta}\right)^2e^{v_0\int_{0}^{\zeta}\frac{d\tilde{\zeta}}{D_e(n_0(\tilde{\zeta}))}}d\zeta}.
\end{equation}

\noindent Equation~(\ref{h_equation}) is the main result of this subsection. 

The first three terms on the right hand side of Eq.~(\ref{h_equation}) represent the standard KPZ equation of surface growth~\cite{kpz, spohn:exact_kpz, kardar_drossel:KPZ_DP2_binary_alloys}. The first term accounts for the growth of a flat surface. The second term accounts for the effects of front tilt: Displacement of a tilted interface along its normal by~$dl$ increases the height by~$dl\left(1+\frac{1}{2}\left(\frac{\partial h}{\partial x}\right)^2\right)$. The third term accounts for the lateral diffusion in the context of surface-deposition models or for the effect of front curvature on the expansion velocity in the context of reaction-diffusion waves. A convex front expands more slowly compared to a flat front because it needs to cover a large area in order to advance by the same distance.

The remaining term on the right hand side of Eq.~(\ref{h_equation}) reflects the differences in the chiralities of the strains. It explicitly breaks the reflection symmetry~$x\to-x$ and, therefore, must be absent in a model of competition between non-chiral strains. The main effect of this new term is to promote growth at the boundaries where the strains move towards each other and suppress growth at the boundaries where the strains move away from each other. To see this, let~$f$ denote the fraction of the strain that has a stronger left-moving bias. With this choice,~$A^{(1)}>A^{(2)}$~(see Eq.~(\ref{flux_decomposition})), and, therefore,~$\alpha>0$. The extra growth due to~$\alpha\frac{\partial f}{\partial x}>0$ then occurs only when the fraction of the left-moving strain increases with~$x$, i.e. when the left-moving strain is to the right of the right-moving strain. For this arrangement, the bias of the strains brings them towards each other, which is precisely the condition for extra growth stated above. We examine the effects of this extra growth quantitatively in a separate section below.

\subsection*{Equation for front composition,~$\boldsymbol{f(t,x)}$}
The equation for~$f(t,x)$ follows from Eq.~(\ref{f_dynamics}) for~$f(t,x,y)$. Note that we denote these two quantities by the same symbol and distinguish them by specifying their arguments explicitly. In the following, we use exclusively~$f(t,x,y)$ until we derive the effective equation for the dynamics of the species fractions at the front.

We begin by changing variables from~$t$,~$x$,~and~$y$ to $\tau$,~$\eta$,~and~$\zeta$. The variable~$\zeta$ denotes the distance between~$(x,y)$ and~$h(t,x)$ as before. The variable~$\eta$ specifies the position along the front and is defined as a curvilinear coordinate orthogonal to~$\zeta$. The variable~$\tau$ equals to~$t$. We introduce this new time variable explicitly to emphasize the fact that~$\frac{\partial}{\partial t}\ne\frac{\partial}{\partial \tau}$ because the change of variables depends on time through~$h(t,x)$.

In the new coordinate system, Eq.~(\ref{f_dynamics}) takes the following form

\begin{equation}
\label{f_dynamics_comoving}
\begin{aligned}
\frac{\partial f}{\partial \tau} = & \frac{\partial}{\partial \zeta}\left(D(n)\frac{\partial f}{\partial \zeta}\right) + \left[ \frac{2D(n)}{n} + S(n)\right]\frac{d n}{d\zeta}\frac{\partial f}{\partial\zeta} + v_0 \frac{\partial f}{\partial\zeta}  \\ 
&+D(n)\frac{\partial^2 f}{\partial \eta^2} - \left[A^{(1)} - A^{(2)}\right] \left[\frac{A^{(1)}}{A^{(1)} - A^{(2)}} - f\right] \frac{d n}{d\zeta}\frac{\partial f}{\partial\eta},
\end{aligned}
\end{equation}

\noindent where we approximated~$\frac{\partial \zeta}{\partial\tau}$ by~$v_0$ to the first order in~$A^{(1)} - A^{(2)}$ and assumed, as before, that~$n$ depends only on~$\zeta$.

To proceed, we assume a time-scale separation between the fast dynamics along~$\zeta$ that quickly equilibrate the values of~$f$ across the thickness of the front and the slow dynamics along the~$\eta$ direction. The fast dynamics are given by the first three terms in Eq.~(\ref{f_dynamics_comoving}) and can be expressed as

\begin{equation}
\label{f_fast}
\left.\left(\frac{\partial f}{\partial \tau}\right)\right|_{\mathrm{fast}} =  L_{\zeta} f,
\end{equation}

\noindent where~$L_{\zeta}$ is a linear operator that is given by

\begin{equation}
\label{f_fast}
L_{\zeta} =  \frac{\partial}{\partial \zeta}\left(D(n)\frac{\partial }{\partial \zeta}\right) + \left[ \frac{2D(n)}{n} + S(n)\right]\frac{d n}{d\zeta}\frac{\partial }{\partial\zeta} + v_0 \frac{\partial }{\partial\zeta}.
\end{equation}

It is clear that~$f=\mathrm{const}$ is an eigenfunction of~$L_{\zeta}$ with a zero eigenvalue. Therefore, the solution of Eq.~(\ref{f_fast}) approaches~$f=\mathrm{const}$ in the long-time limit. The value of this constant can obtained by multiplying Eq.~(\ref{f_fast}) by~$l_0(\zeta)$, the left eigenfunction of~$L_{\zeta}$ with zero eigenvalue, and then integrating over~$\zeta$. After integrating by parts, the right hand side vanishes, so~$\frac{\partial}{\partial \tau}\int_{-\infty}^{+\infty}l_0fd\zeta=0$, and the projection of~$f$ on~$l_0$ is conserved. The value of this left eigenfunction can be easily obtained by the standard methods~\cite{birzu:semi_pushed, matin:cooperative_migration} and is given by\footnote{This left eigenfunction is related to the probability that a mutant arises at position~$\zeta$ and reaches fixation at the front some time afterwards~\cite{hallatschek:diversity_wave, birzu:semi_pushed, matin:cooperative_migration}.}

\begin{equation}
\label{l_0}
l_0 = n^2(\zeta)e^{\int_{0}^{\zeta}\frac{v_0+S(n(\tilde{\zeta}))}{D(n(\tilde{\zeta}))}d\zeta}.
\end{equation}

We can now eliminate the fast dynamics by projecting both sides of Eq.~(\ref{f_dynamics_comoving}) on~$l_0(\zeta)$. This procedure provides a natural definition of the coarse-grained variable~$f(t,\eta)$ in terms of~$f(t,\zeta,\eta)$

\begin{equation}
\label{ftx_definition}
f(t,\eta) = \frac{\int_{-\infty}^{+\infty}l_0(\zeta)f(\tau,\zeta,\eta)d\zeta}{\int_{-\infty}^{+\infty}l_0(\zeta)d\zeta},
\end{equation}

\noindent and leads to the following equation for~$f(t,\eta)$

\begin{equation}
\label{f_equation_eta}
\frac{\partial }{\partial \tau} f(t,\eta) = D_f\frac{\partial^2 }{\partial \eta^2} f(t,\eta)  + \beta [f^* - f(t,\eta)] \frac{\partial}{\partial\eta}f(t,\eta),
\end{equation}

\noindent where

\begin{equation}
\label{D_f}
D_f =  \frac{\int_{-\infty}^{+\infty}l_0(\zeta)D(n(\zeta))d\zeta}{\int_{-\infty}^{+\infty}l_0(\zeta)d\zeta},
\end{equation}

\begin{equation}
\label{beta}
\beta =  - \frac{\int_{-\infty}^{+\infty}l_0(\zeta)[A^{(1)}(n(\zeta)) - A^{(2)}(n(\zeta))]\frac{dn}{d\zeta}d\zeta}{\int_{-\infty}^{+\infty}l_0(\zeta)d\zeta},
\end{equation}

and

\begin{equation}
\label{f*}
f^* =  \frac{\int_{-\infty}^{+\infty}l_0(\zeta)A^{(1)}(n(\zeta))\frac{dn}{d\zeta} d\zeta}{\int_{-\infty}^{+\infty}l_0(\zeta)[A^{(1)}(n(\zeta)) - A^{(2)}(n(\zeta))]\frac{dn}{d\zeta}d\zeta}.
\end{equation}

\noindent Note that we assumed that~$f(\tau,\zeta,\eta)\approx f(\tau,\eta)$ in all terms representing the slow dynamics along~$\eta$. 

The final step of the derivation is to change from~$\tau$ and~$\eta$ back to~$t$ and~$x$. To the first order in~$\frac{\partial h}{\partial x}$, we find that~$\frac{\partial f}{\partial \eta} = \frac{\partial f}{\partial x}$ and~$\frac{\partial f}{\partial \tau} = \frac{\partial f}{\partial t} + \frac{\partial f}{\partial x}\frac{\partial x}{\partial \tau }= \frac{\partial f}{\partial t} + v_0 \frac{\partial f}{\partial x}\frac{\partial h}{\partial x}$. The last equality follows from the geometrical fact that a stationary point on a front tilted by a small angle~$\frac{\partial h}{\partial x}$ moves with velocity~$-v_0\frac{\partial h}{\partial x}$ with respect to the~$x$-axis. Upon performing these substitutions, Eq.~(\ref{f_equation_eta}) takes the following form

\begin{equation}
\label{f_equation}
\frac{\partial f }{\partial t}  = D_f\frac{\partial^2 f}{\partial x^2} + \beta (f^* - f) \frac{\partial f}{\partial x} +  v_0 \frac{\partial h}{\partial x}\frac{\partial f}{\partial x}.
\end{equation}

Equation~(\ref{f_equation}) is the main result of this subsection. The first term of this equation describes the diffusive intermixing of the strains. The second term accounts for the movement of strains due to chirality. The third term describes the ``kinematic'' coupling between front shape and front composition and accounts for the sliding of a tilted front relative to the~$x$-axis. In other words, the species fractions at a given~$x$ change simply because the same~$x$ corresponds to different locations on the front~(specified by~$\eta$) at different times. Without the last term, Eq.~(\ref{f_equation}) is the viscous Burgers' equation that describes dissipative flow of conserved quantities in one spatial dimension and is used as a toy model of fluid dynamics and traffic flow~\cite{bateman:burgers_first, burgers:original, sachdev:diffusive_waves, whitham:waves}. We find that the behavior of the boundaries is largely captured by the Burgers' equation except the last term in~Eq.~(\ref{f_equation}) introduces an extra drift down the slopes of~$h(t,x)$. The dynamics of the boundaries is discussed in a separate section below.

The primary results of this section are Equations~(\ref{f_equation}) and~(\ref{h_equation}) which couples the shape of the front with population dynamics occuring at the front. Such phenomenological models have been proposed to study systems such as the surface growth of binary alloy films~\cite{kardar_drossel:KPZ_DP2_binary_alloys}

\section{Derivation of the effective theory from phenomenological considerations}

This section derives the effective theory stated by Eqs.~(\ref{h_equation}) and~(\ref{f_equation}) from phenomenological considerations. Specifically, we perform a gradient expansion assuming that the spatial variations are slow and consider the most general set of equations for~$h(t,x)$ and~$f(t,x)$ that contain terms with up to two spatial derivatives:

\begin{equation}
\label{h_general}
\frac{\partial h }{\partial t} = H_{0} + H_{h_x}\frac{\partial h }{\partial x} + H_{f_x}\frac{\partial f }{\partial x} + H_{h_{xx}}\frac{\partial^2 h }{\partial x^2} + H_{(h_x)^2}\left(\frac{\partial h }{\partial x}\right)^2 + H_{h_xf_x}\frac{\partial h}{\partial x}\frac{\partial f }{\partial x} + H_{(f_x)^2}\left(\frac{\partial f }{\partial x}\right)^2 + H_{f_{xx}}\frac{\partial^2 f }{\partial x^2}, 
\end{equation}

\noindent and

\begin{equation}
\label{f_general}
\frac{\partial f }{\partial t} = F_{0} + F_{h_x}\frac{\partial h }{\partial x} + F_{f_x}\frac{\partial f }{\partial x} + F_{h_{xx}}\frac{\partial^2 h }{\partial x^2} + F_{(h_x)^2}\left(\frac{\partial h }{\partial x}\right)^2 + F_{h_xf_x}\frac{\partial h}{\partial x}\frac{\partial f }{\partial x} + F_{(f_x)^2}\left(\frac{\partial f }{\partial x}\right)^2 + F_{f_{xx}}\frac{\partial^2 f }{\partial x^2}, 
\end{equation}

\noindent where every coefficient could depend on~$f$, but not on~$h$ because the space is invariant under translations.

To understand the role and origin of each term, it is important to know the effects of the following transformations: reflection in the~$y$-axis~($x\to-x$), exchange of strains labels, and rotation of the front by a small angle~$\phi$. Let us examine each transformation in more detail. 
Reflection in the~$y$-axis changes the sign of~$x$ and the signs of the chiralities of the strains. By chiralities, we do not necessarily mean the ones defined in Eq.~(\ref{flux_decomposition}), but any two numbers that capture the chiral nature of the two strains such that the chirality is zero for non-chiral strains, positive for left-moving strains, and negative for right-moving strains. All terms that appear in Eqs.~(\ref{h_general}) and~(\ref{f_general}) due to chirality can be expanded in Taylor series in powers of the difference between the chiralities of the two strains~$\Delta A = A^{(1)}-A^{(2)}$. The constant term in the series must vanish because the morphology of a compact microbial colony is the same for any two strains with equal chiralities regardless of the magnitude of this chirality. In addition to~$\Delta A$, the terms could depend on~$f^*=A^{(1)}/(A^{(1)}-A^{(2)})$, which does not change when both~$A^{(1)}$ and~$A^{(2)}$ change sign. These considerations and the fact that reflection in the~$y$ should not alter the nature of the competition leads us to the following conclusion: Eqs.~(\ref{h_general}) and~(\ref{f_general}) must remain invariant under~$x\to-x$ and~$\Delta A\to-\Delta A$.

The model formulation should also be invariant under the exchange of species labels. This exchange results in the following transformations~$f\to1-f$,~$\Delta A\to -\Delta A$, and~$f^*\to1-f^*$, which impose another symmetry requirement on Eqs.~(\ref{h_general}) and~(\ref{f_general}).

Rotation by a small angle~$\phi$ leads to a change of variable from~$x$ and~$h$ to~$\tilde{x} = x/\cos\phi$ and~$\tilde{h}=h\cos\phi-x\sin\phi$. When space is isotropic this transformation should also leave Eqs.~(\ref{h_general}) and~(\ref{f_general}) unchanged. However, because these equations assume that~$\frac{\partial h}{\partial x}$ is small, we only require invariance up to the second order in~$\phi$. Note that, this symmetry does not hold for non-isotropic spaces, which appear, for example, in lattice-based simulations. In the equation for~$f(t,x)$ a rotation by angle~$\phi$ also modifies~$\frac{\partial f}{\partial t}$ to~$\frac{\partial f}{\partial t}-v_0\frac{\partial f}{\partial \tilde{x}}\tan\phi$.

In addition to the above symmetries, we must also require that strains cannot be spontaneously created. This requirement implies that all of the terms in Eq.~(\ref{f_equation}) should vanish when~$f\to0$ and~$f\to1$.  

We now analyze each of the terms in Eqs.~(\ref{h_general}) and~(\ref{f_general}) separately and then discuss the implications of these results for the form of the effective model of the competition at the front. 

\paragraph{Term~$H_0 + H_{(h_x)^2}\left(\frac{\partial h }{\partial x}\right)^2$}. These two terms are considered together because rotational symmetry requires that~$H_{(h_x)^2}=H_0/2$. When~$f$ and~$h$ do not depend on~$x$,~$H_0$ is the only term contributing to~$\frac{\partial h}{\partial t}$, so we can interpret it as the front velocity~$v_0$. This velocity could depend on~$f$, for example, when species are cross-feeding or one of the species is more fit. Even when chirality is the only difference between the species,~$v_0$ could still be~$f$-dependent, but this dependence should scale as~$(\Delta A)^2$ because the preceding term in the Taylor expansion must vanish due to mirror symmetry\footnote{If the two strains have identical expansion velocities when grown by themselves, then the correction to~$v_0$ should scale as~$f(1-f)(\Delta A)^2$ to the leading order in~$\Delta A$.}. Thus,~$v_0=\mathrm{const}$ is a reasonable approximation for small differences in strain chiralities. We also note that in lattice-based models, which explicitly violate rotational symmetry,~$H_{(h_x)^2}=H_0/2$ does not have to hold. The deviations from this equality indicate how the expansion velocity depends on the orientation of the front relative to the lattice.

\paragraph{Term~$H_{h_{xx}}\frac{\partial^2 h }{\partial x^2}$}. For~$H_{h_{xx}}=\mathrm{const}$, this term corresponds to the diffusion term in Eq.~(\ref{h_equation}). The dependence on~$f$ is possible, but should scale as~$(\Delta A)^2$ for the same reasons as above. The three terms discussed so far constitute the standard KPZ equation without noise~\cite{spohn:exact_kpz}.

\paragraph{Term~$H_{f_x}\frac{\partial f }{\partial x}$}. This term corresponds to~$\alpha\frac{\partial f}{\partial x}$ in Eq.~(\ref{h_equation}) and is the only new term that appeared due to chirality in our derivation of the effective model via the moving-boundary approximation. Because~$\frac{\partial f}{\partial x}$ changes sign under mirror symmetry,~$H_{f_x}$ should be odd in~$\Delta A$. Our expression for~$\alpha$ in Eq.~(\ref{alpha}) can therefore be viewed as the first term in the Taylor expansion of~$H_{f_x}$ in powers of~$\Delta A$. The symmetry under the exchange of the labels further implies that~$\alpha$ can not depend on~$f$ linearly, so our result that~$\alpha=\mathrm{const}$ should be a reasonable approximation to a more general model. In the reaction-diffuson model, the dependence of~$\alpha$ on~$f$ can arise from the dependence of~$A^{(\alpha)}$ on~$f$.

\paragraph{Term~$H_{h_x}\frac{\partial h }{\partial x}$}. This term violates rotational symmetry and, therefore, must be absent in a continuum model, but it should appear in a lattice-bades model. Because~$\frac{\partial h }{\partial x}$ changes sign under the mirror symmetry,~$H_{h_x}\sim\Delta A$ at the leading order. The label exchange symmetry imposes a further constraint\footnote{In the simplest case, the relative chirality could be quantified by a single~$f^*$ entering all relevant terms. However, the possible dependence of~$A^{(\alpha)}$ on~$n$ together with Eq.~(\ref{f*}) suggest that slightly different~$f^*$ might appear in different terms. To account for such a possibility, we added an index to~$f^*$ that signifies the term in the equation that this particular~$f^*$ is associated with. For simplicity, we omit such indexes in the rest of this section.}~$H_{h_x}\approx\gamma(f_{\gamma}^*-f)$, where~$\gamma\sim\Delta A$. Indeed,~$H_{h_x}$ simply equal to $\gamma$ without the~$f$-dependence changes sign under the label exchange and thus does not leave the equation invariant.

\paragraph{Term~$H_{h_xf_x}\frac{\partial h}{\partial x}\frac{\partial f }{\partial x}$}. Similar to the term just discussed, this term violates the rotational symmetry and, therefore, should be excluded. In lattice-based models,~$H_{h_xf_x}\frac{\partial h}{\partial x}\frac{\partial f }{\partial x}$ is allowed, but we expect that its contribution is subleading to that from~$H_{h_x}\frac{\partial h }{\partial x}$ because of the extra spatial derivative, which is small under our assumption of slow spatial variation. The mirror symmetry requires that~$H_{h_xf_x}\sim(\Delta A)^2$.

\paragraph{Term~$H_{(f_x)^2}\left(\frac{\partial f }{\partial x}\right)^2$}. Since~$\left(\frac{\partial f }{\partial x}\right)^2$ is invariant under~$x\to-x$, this term can describe competition between non-chiral strains. One possible origin of this term in the reaction-diffusion model is the dependence of~$D^{(1)} - D^{(2)}$ on~$f$. When~$H_{(f_x)^2}\left(\frac{\partial f }{\partial x}\right)^2$ arises due to chirality differences, we expect that~$H_{(f_x)^2}\sim(\Delta A)^2$ due to the mirror symmetry.

\paragraph{Term~$H_{f_{xx}}\frac{\partial^2 f }{\partial x^2}$}. Similar to the term just discussed, $H_{f_{xx}}\frac{\partial^2 f }{\partial x^2}$~could appear in a model without any chirality difference, for example, due to unequal dispersal coefficients of the two strains~($D^{(1)}\ne D^{(2)}$). The reflection symmetry and the symmetry due to the exchange of labels suggest that~$H_{f_{xx}}$ should scale as~$\Delta A(f^*-f)$ to the lowest order in~$\Delta A$ assuming the simplest dependence on~$f$. More complex dependence is also possible, for example,~$F_{h_x}=\Delta A (f^*-f)Q(f(1-f))$, where~$Q$ is an arbitrary function. 

This completes the analysis of the terms in Eq.~(\ref{h_general}), so we proceed with the same analysis for Eq.~(\ref{f_general}).

\paragraph{Term~$F_{0}$}. This term arises due to the difference in the growth rates of the strains,~$g^{(\alpha)}$, and describes the effects of natural selection such as exclusion, coexistence, or bistability. Because we are interested in the effects of chirality, we only consider strains with the same fitness, for which~$F_0=0$.

\paragraph{Term~$F_{h_x}\frac{\partial h }{\partial x}$}. Since this term violates rotational symmetry, it requires that the space is not isotropic, for example, due to using a lattice-based model to simulate population dynamics. The mirror symmetry requires that~$F_{h_x}$ is proportional to~$\Delta A$ at the lowest order in the difference between the chiralities. Further,~$F_{h_x}$ must vanish when~$f=0$ or~$f=1$ because the strains cannot interconvert between each other. Thus, the simplest form of this term is~$F_{h_x}\sim\Delta A f(1-f)$, but a more complicated dependence on~$f$ is possible similar to the case of~$H_{f_{xx}}\frac{\partial^2 f }{\partial x^2}$ discussed above. One potentially important effect due to~$F_{h_x}$ is the induced selection on the sides of the bulges formed around in-flow boundaries. When~$F_{h_x}>0$, selection favors the invasion of either side of the bulge by the strain dominating the other side. Hence, positive~$F_{h_x}$ promotes strain intermixing. Negative~$F_{h_x}$ opposes strain intermixing because selection suppresses the growth of the strain dominating the opposite side of the bulge.  

\paragraph{Term~$F_{f_x}\frac{\partial f }{\partial x}$}. This term is equivalent to~$\beta(f^*-f)\frac{\partial f }{\partial x}$ in Eq.~(\ref{f_equation}). To satisfy the requirements of the mirror and label exchange symmetries,~$F_{f_x}$ should scale as~$\Delta A(f^*-f)$ consistent with the results from the reaction-diffusion model. More general dependence on~$f$ is also possible; for example~$F_{f_x}$ could scale as~$\Delta A(f^*-f)Q(f(1-f))$, where~$Q$ is an arbitrary function. Thus,~$\beta$ could depend on~$f$ in Eq.~(\ref{f_equation}). This dependence, however, is unlikely to alter any of the qualitative conclusions because the main aspects of the strain motion due to chirality are already captured by constant~$\beta$. In particular, the dependence of~$\beta$ on~$f$ contributes only to the width of the strain boundaries and is irrelevant in the domains occupied by a single strain, where $f=\mathrm{const}$. At the boundaries, however, the phenomenological model does not account for other important factors such as stochasticity and large spatial gradients, so neglecting the dependence of~$\beta$ on~$f$ is a reasonable simplification.  

\paragraph{Term~$F_{h_{xx}}\frac{\partial^2 h }{\partial x^2}$}. Since~$\frac{\partial^2 h }{\partial x^2}$ is invariant under~$x\to-x$, this term could occur even without chirality differences. When it occurs due to chirality differences, the symmetries require that~$F_{h_{xx}}\sim(\Delta A)^2(f^*-f)$. More importantly, a term without a gradient of~$f$ cannot arise in our reaction-diffusion model because~$f=\mathrm{const}$ is a solution of Eq.~(\ref{f_dynamics}) for any~$h(t,x)$. Therefore, the origin of~$F_{h_{xx}}\frac{\partial^2 h }{\partial x^2}$ must be related to the difference in a fitness component of strains that couples the relative growth rate and the local front curvature.  

\paragraph{Term~$F_{(h_x)^2}\left(\frac{\partial h }{\partial x}\right)^2$}. All of the comments that we made about~$F_{h_{xx}}\frac{\partial^2 h }{\partial x^2}$ apply to this term as well. In addition,~$\left(\frac{\partial h }{\partial x}\right)^2$ is not invariant under rotations, so this term can arise only when rotational symmetry is broken, for example, due to lattice effects.  

\paragraph{Term~$F_{h_xf_x}\frac{\partial h}{\partial x}\frac{\partial f }{\partial x}$}. To satisfy rotational invariance, we must set~$F_{h_xf_x}$ to~$v_0$. Lattice effects however can result in a violation of this constraint and make the boundary between two strains move with different velocities~(relative to the front) for different front orientations with respect to the lattice.

\paragraph{Term~$F_{(f_x)^2}\left(\frac{\partial f }{\partial x}\right)^2$}. Since~$\frac{\partial^2 h }{\partial x^2}$ is invariant under~$x\to-x$, this term could occur even without chirality differences. When it occurs due to chirality differences, the mirror and label symmetries require that~$F_{(f_x)^2}\sim(\Delta A)^2(f^*-f)$, i.e. this term is of higher order in the difference between the chiralities. In addition, this term violates another symmetry: Galilean invariance. Because the term is independent of~$h$, we can consider a flat front containing left and right biased strains. Since one can always choose a reference frame moving along the front in which the biases are equal and opposite, $F_{(f_x)^2}$~cannot depend on~$f^*$, which is not invariant under this transformation. The presence of~$f^*$ is however required by the symmetry under the exchange of the labels. Therefore,~$F_{(f_x)^2}\left(\frac{\partial f }{\partial x}\right)^2$ can appear only in models that contain a preferred reference frame and thereby explicitly break Galilean invariance. This situation arises naturally in lattice-based models because their formulation often depends on the lattice being at rest.    

\paragraph{Term~$F_{f_{xx}}\frac{\partial^2 f }{\partial x^2}$}. This term corresponds to~$D_f\frac{\partial^2 f }{\partial x^2}$ in Eq.~(\ref{f_equation}), which describes the diffusive intermixing of the strains. The diffusion coefficient could in principle depend on~$f$, but the mirror symmetry requires that this dependence is at least quadratic in~$\Delta A$. Therefore, in the following, we assume that~$D_f=\mathrm{const}$ consistent with the results from the reaction-diffusion model~(Eq.~\ref{D_f}). 

After analyzing all of the terms in Eqs.~(\ref{h_general}) and~(\ref{f_general}), we conclude that, to the leading order in~$\Delta A$, Eqs.~(\ref{h_equation}) and~(\ref{f_equation}) capture all possible contributions to the evolution of~$h(t,x)$ and~$f(t,x)$ consistent with the symmetries of the problem. Therefore,~Eqs.~(\ref{h_equation}) and~(\ref{f_equation}) should hold even when some of the assumptions underlying their derivation from the reaction-diffusion model are violated. The phenomenological description further provides a clear path to accounting for the lattice effects in simulations and to including the effects that higher order in~$\Delta A$. 

\section{Deterministic behavior of in-flow and out-flow boundaries in flat fronts}

This section considers the dynamics of isolated domain boundaries located in flat regions of the front without tilt~($\frac{\partial h}{\partial x}=0$). We determine the velocity of in-flow and out-flow boundaries and explain why these two types of boundaries have different width. Throughout this section we neglect fluctuations in~$f$ due to genetic drift. Demographic fluctuations, however, play a major role in our simulations and microbial experiments~\cite{korolev:review, korolev:amnat}, so the discussion of boundary shape is included largely to provide intuitive understanding rather than quantitative description of the dynamics. In the limit of large carrying capacity,~$N$, our simulations do reproduce the behavior described below, but for moderate~$N$ genetic drift leads to a qualitative change in the dynamics. Specifically, strong demographic fluctuations limit the spreading of domain boundaries and restrict strain intermixing to a finite region~\cite{hallatschek:noisy_fisher}. The deterministic dynamics discussed below influence the size of this region and make in-flow boundaries wider than out-flow boundaries.

\subsection*{Boundary velocity}
By setting~$\frac{\partial h}{\partial x}$ to zero in Eq.~(\ref{f_equation}), we obtain
 
\begin{equation}
\label{f_burgers}
\frac{\partial f }{\partial t}  = D_f\frac{\partial^2 f}{\partial x^2} + \beta (f^* - f) \frac{\partial f}{\partial x},
\end{equation}

\noindent which becomes equivalent to the viscous Burgers' equation upon identifying~$f-f^*$ with the flow velocity~\cite{sachdev:diffusive_waves, whitham:waves}. The Burgers' equation has been extensively studied, so the properties of its solutions are well-characterized~\cite{burgers:original, sachdev:diffusive_waves, whitham:waves}. The main purpose of this section is to provide a short summary of the standard results in the language of our model.

The analysis of Eq.~(\ref{f_burgers}) simplifies upon changing into a reference frame moving with velocity 

\begin{equation}
v_{\parallel} = (\frac{1}{2}-f^*)\beta.
\label{v_parallel}
\end{equation}

\noindent Indeed, by letting~$\mathfrak{z}=x-v_{\parallel}t$, we find that Eq.~(\ref{f_burgers}) takes the following form

\begin{equation}
\label{f_burgers_comoving}
\frac{\partial f }{\partial t}  = D_f\frac{\partial^2 f}{\partial \mathfrak{z}^2} + \beta \left(\frac{1}{2} - f\right) \frac{\partial f}{\partial \mathfrak{z}},
\end{equation}

\noindent which is the same as Eq.~(\ref{f_burgers}) but with~$f^*$ replaced by~$\frac{1}{2}$. Because~$f^*=\frac{1}{2}$ corresponds to exactly opposite chiralities~(see Eq.~(\ref{f*})), the dynamics in the reference frame moving with velocity~$v_{\parallel}$ are equivalent to those of two oppositely chiral strains. From the symmetry considerations, it follows that the boundaries between the two oppositely chiral strains must remain stationary. Therefore,~$v_{\parallel}$ defined by Eq.~(\ref{v_parallel}) is the velocity of the boundaries in the original reference frame.\footnote{This result assumes that~$f$ is either~$0$ or~$1$ on both sides of the boundary, as it is in our simulations. The more general result is that~$v_{\parallel}=\beta\left(\frac{f_++f_-}{2}-f^*\right)$, where~$f_-$ and~$f_+$ are the values of~$f$ on the two sides of the boundary.} 

Equation~(\ref{v_parallel}) can be obtained more directly. The first step is to observe that Eq.~(\ref{f_burgers}) conserves the integral of~$f$ over~$dx$ because~$-\frac{\partial f}{\partial t}$ equals the divergence of the flux~$J=-D_f\frac{\partial f}{\partial x} + \frac{\beta}{2} (f - f^*)^2$. The second step is to find a reference frame where the fluxes in and out of the domain boundary balance. Due to the conservation of~$\int fdx$, the boundary must remain stationary when there is no net flux, so the boundary velocity is given by the velocity of the reference frame. This approach yields the same value of~$v_{\parallel}$ as in Eq.~(\ref{v_parallel}).

\subsection*{Shape of boundaries}
We now consider the shapes of in-flow and out-flow boundaries in the co-moving reference frame. For an out-flow boundary, Eq.~(\ref{f_burgers_comoving}) admits the following stationary solution

\begin{equation}
\label{out_flow_shape}
f(\mathfrak{z}) = \frac{1}{2}\left[1-\tanh\left(\frac{\beta}{4D_f}\mathfrak{z}\right)\right].
\end{equation}

\noindent Thus, out-flow boundaries reach a finite width on the order of~$D_f/\beta$ even in the absense of noise. The out-flouw boundaries become infinitely sharp in the limit of~$D_f\to0$, which corresponds to the inviscid Burgers' equation. Such sharp changes in~$f$ are termed shocks in the language of nonlinear partical differential equations. 

The behavior of in-flow boundaries is completely different. Instead of reaching a fixed shape, they widen indefinitely. The temporal evolution of the boundary shape can be determined by solving Eq.~(\ref{f_burgers_comoving}) equation exactly. This is done by the following Cole-Hopf transformation:

\begin{equation}
f = \frac{1}{2} + \frac{2D_f}{\beta}\frac{\partial \ln w}{\partial \mathfrak{z}},
\label{ch_burgers}
\end{equation}

\noindent which yields the standard diffusion equation for~$w$:

\begin{equation}
\label{w_ch}
\frac{\partial w }{\partial t}  = D_f\frac{\partial^2 w}{\partial \mathfrak{z}^2}.
\end{equation}

The qualitative behavior of in-flow boundaries can also be obtained in a simpler way by neglecting the effect of diffusion; this is a very good approximation on intermediate time scales~\cite{whitham:waves}. Setting~$D_f$ to zero reduces Eq.~(\ref{f_burgers}) to a first order partial differential equation, which can be solved by the method of characteristics. For a step-function initial condition~($f(0,x)=\theta(x)$), the solution reads 

\begin{equation}
\label{in_flow_shape}
f(t,x) = \left\{
		\begin{aligned}
		& 1\;\; x>\beta(1-f^*)t,\\
		& f^* + \frac{x}{\beta t}\;\; x\in[-\beta f^* t,\; \beta(1-f^*)t],\\
		& 0\;\; x < -\beta f^* t.
		\end{aligned}
		\right.
\end{equation}

\noindent Thus, the width of the boundary grows as~$\beta t$. While this widening is arrested by strong genetic drift, we nevertheless expect in-flow boundaries to be wider than out-flow boundaries. Our simulations agree with this expectation; see Figs.2 and 3 in the main text.

\subsection*{Interaction between boundaries in the Burgers' equation}
So far, we considered the dynamics of isolated boundaries, which is a good description for times shorter than the time required for a widening in-flow boundary to reach the nearest out-flow boundary. Beyond this time scale, the boundaries begin to interact. This interaction is largely irrelevant in the context of microbial colonies because strong genetic drift arrests the widening of out-flow boundaries; as a result, the boundaries come in contact only via a random walk or due to the bulge-induced motion. 

For completeness, we briefly destribe the effect of boundary interactions in the context of the Burgers' equation, but we emphasize that these dynamics do not occur in the context of competition between chiral strains in growing colonies. Burgers' equation predicts that boundary interaction induces the motion of out-flow boundaries in the opposite direction compared to results of our simulations and the prediction based on the bulge-induced motion of out-flow boundaries. The interaction between boundaries also reduces spatial variations in~$f(t,x)$ until the stationary solution~$f(t,x)=\mathrm{const}$ is reached. The deviations from~$f(t,x)=\mathrm{const}$ decay as~$t^{-1/2}$ for a pair of isolated boundaries and as~$t^{-1}$ for a periodic array of boundaries~\cite{whitham:waves}.

\section{Deterministic solution for bulge and dip shape near a sharp boundary}

In this section, we obtain the shape of bulges and dips at in-flow and out-flow boundaries in the effective theory given by Eqs.~(\ref{h_equation}) and~(\ref{f_equation}). In general, this system of coupled equations is difficult to solve, so we resort to two additional approximations. The first approximation is that we decouple the equations and solve for~$h(t,x)$ by assuming that~$f(t,x)$ is known. The second approximation is that we consider the limit of a very sharp boundary and assume that~$f(t,x)=\theta(\pm(x-v_{\parallel}t))$. Here,~$+$~refers to in-flow and~$-$~to out-flow boundaries respectively; $\theta(x)$ is the Heaviside step function; and~$v_{\parallel}$ is the velocity of the boundary. This assumption is justified in the limit of strong genetic drift, which prevents strain intermixing through local stochastic exctinctions of one of the strains. In microbial colonies, genetic drift is typically very strong and one does indeed observe very narrow boundaries~\cite{korolev:amnat,korolev:review,hallatschek:sectors}. The sharp boundary approximation can also be viewed as a simplifying assumption that preserves the qualitative aspects of the problem by neglecting the effects due to an additional length scale associated with boundary width. 

The velocity of the boundary depends on~$f^*$. To the first approximation in~$\Delta A$, it is natural to neglect the contribution of the front shape near the boundary and assume that~$v_{\parallel}$ is given by its value for a flat front in Eq.~(\ref{v_parallel}). This assumption, however, needs a careful consideration because we show in this and following sections that the tilt of the front due to a bulge is proportional to~$\Delta A$ and that this tilt induces a velocity linear in~$\Delta A$ for a boundary trapped on the slope of the bulge. Because a boundary associated with a bulge or a dip is positioned between two slopes tilted in opposite directions, the induced velocities could effectively cancel\footnote{One needs to analyze the deviations from the linear shape of the bulge to make this argument precise. In particular, no modification of~$v_{\parallel}$ is expected if the boundary is located on the flat portion of the bulge's top where~$\frac{\partial h}{\partial x}=0$; see Eq.~(\ref{boundary_motion}).} at least to the first order in~$\Delta A$. For a special case of~$f^*=\frac{1}{2}$, one can however argue that~$v_{\parallel}=0$ purely from the symmetry considerations. Below, we analyze this simpler case first and then generalize the result for~$v_{\parallel}\ne0$. Our results in this section do not depend on the specific expression for the boundary velocity, so we treat~$v_{\parallel}$ as an arbitrary parameter to preserve generality.

\subsection*{Bulge shape at a stationary in-flow boundary,~$\boldsymbol{f^*}=\frac{1}{2}$}
Upon substituting~$f(t,x)=\theta(x)$ into Eq.~(\ref{h_equation}), we obtain

\begin{equation}
\label{symmetric_bulge_formulation}
\frac{\partial h}{\partial t} = v_0 + \frac{v_0}{2}\left(\frac{\partial h}{\partial x}\right)^2 + D_h \frac{\partial^2 h}{\partial x^2} + \alpha\delta(x),
\end{equation}

\noindent where~$\delta(x)$ is the Dirac delta function. A reasonable choice of the initial conditions is~$h(0,x)=0$.

Before describing the formal solution, let us observe that the long-time limit of~$h(t,x)$ can be immediately guessed from Eq.~(\ref{symmetric_bulge_formulation}). Indeed, the first term contributes an additive term to the solution equal to~$v_0t$. The Delta function leads a discontinuity in~$\frac{\partial h}{\partial x}$ at~$x=0$ and thereby induces front tilt, which increases the front velocity from~$v_0$ to a higher value through~$\frac{v_0}{2}\left(\frac{\partial h}{\partial x}\right)^2$~term. The diffusion term is exactly zero provided~$h(t,x)$ is a linear function of~$x$. Thus, the solution should be a linear combination of a term proportional to time~$t$ and a term proportional to the absolute value of the position~$|x|$. Direct substitution of this ansatz reveals that

\begin{equation}
\label{symmetric_bulge_asymptotic_solution}
h(t,x)_{\mathrm{bulge}} = v_0\left(1+\frac{\alpha^2}{8D_h^2}\right)t-\frac{\alpha}{2D_h}|x|.
\end{equation}

We now proceed to derive this result more formally by solving Eq.~(\ref{symmetric_bulge_formulation}) for all~$t$. The first step is based on the Cole-Hopf transformation~\cite{whitham:waves}:

\begin{equation}
\label{u}
h = v_0t + \frac{2D_h}{v_0}\ln u,
\end{equation}

\noindent which leads to a linear equation for the new dynamical variable~$u$ 

\begin{equation}
\label{u_dynamics_symmetric}
\frac{\partial u}{\partial t} =D_h \frac{\partial^2 u}{\partial x^2} + \frac{v_0\alpha}{2D_h}u\delta(x).
\end{equation}

\noindent The initial condition transforms to~$u(0,x)=1$. We note in passing that, for~$\alpha<0$, Eq.~(\ref{u_dynamics_symmetric}) coincides with the equation that describes the decay of neutral diversity in a one-dimensional population~(Eq.~(33) in Ref.~\cite{korolev:review}); therefore, our solution~(Eq.~\ref{u_symmetric_solution}) provides a closed-form expression for the decay of heterozygosity.

To proceed, we replace the term with the delta function by the following boundary conditions

\begin{equation}
\label{bc_u_symmetric}
\begin{aligned}
&\lim_{x\to+0}\frac{\partial u}{\partial x} =-\frac{v_0\alpha}{4D^2_h}u(0),\\
&\lim_{x\to-0}\frac{\partial u}{\partial x} =\frac{v_0\alpha}{4D^2_h}u(0),\\
\end{aligned}
\end{equation}

\noindent which are obtained by integrating Eq.~(\ref{u_dynamics_symmetric}) over~$x$ in the vicinity of~$x=0$, neglecting the contribution form~$\frac{\partial u}{\partial t}$, and using the mirror symmetry of the problem. After this step, we apply the Laplace transform in time and obtain the solution of Eq.~(\ref{u_dynamics_symmetric}) 

\begin{equation}
\label{u_solution_Laplace_symmetric}
u(s,x)=\frac{1}{s}\left(1 + \frac{\alpha v_0}{4D_h^2}\frac{e^{-\sqrt{\frac{s}{D_h}}|x|}}{\sqrt{\frac{s}{D_h}}-\frac{\alpha v_0}{4D_h^2}}\right),
\end{equation}
    
\noindent in terms of~$u(s,x) = \int_{0}^{+\infty}e^{st}u(t,x)dt$.

The inverse Laplace transform then provides the solution for~$u(t,x)$:

\begin{equation}
\label{u_inverse_Laplace_symmetric}
u(t,x)=\frac{1}{2\pi i} \int_{a-i\infty}^{a+i\infty}\frac{e^{st}}{s}\left(1 + \frac{\alpha v_0}{4D_h^2}\frac{e^{-\sqrt{\frac{s}{D_h}}|x|}}{\sqrt{\frac{s}{D_h}}-\frac{\alpha v_0}{4D_h^2}}\right)ds,
\end{equation}

\noindent where~$a$ is a positive number larger than the real part of any singularity of the integrand. 

The integrand in Eq.~\ref{u_inverse_Laplace_symmetric} has a branch cut along the negative~$x$-axis and two poles at~$s=0$ and~$s=\frac{\alpha^2v_0^2}{16D_h^3}$. Therefore, we can simplify the integral as follows

\begin{equation}
\label{u_poles_symmetric}
u(t,x)=\theta(t)\left[1 + 2e^{\frac{v_0}{2D_h}\left( \frac{v_0\alpha^2}{8D_h^2}t-\frac{\alpha}{2D_h}|x| \right)} + \frac{1}{2\pi i} \frac{\alpha v_0}{4D_h^2}\int_{\mathcal{C}}\frac{e^{st}}{s}\frac{e^{-\sqrt{\frac{s}{D_h}}|x|}}{\sqrt{\frac{s}{D_h}}-\frac{\alpha v_0}{4D_h^2}}ds\right],
\end{equation}

\noindent where the contour~$\mathcal{C}$ first goes just under the negative~$x$-axis from~$-\infty-0i$ to~$0$, then around~$s=0$, and finally just above the negative~$x$-axis from~$0$ to~$\infty+0i$. In the following, we consider only~$t>0$ and, therefore, omit~$\theta(t)$ in all formulas. 

It is easy to see that the second term in Eq.~(\ref{u_poles_symmetric}) describes the long-time limit of~$u(t,x)$. Indeed, this term arises due to the pole at~$s=\frac{\alpha^2v_0^2}{16D_h^3}$, which is the dominant singularity because it has the largest real part. The first term in Eq.~(\ref{u_poles_symmetric}) arises due to the initial conditions and the final term describes transient dynamics and the transition from the asymptotic long-time limit at small~$x$ to unperturbed front,~$u=1$, at large~$x$.

The integral in Eq.~(\ref{u_poles_symmetric}), which we denote as~$I$, can be evaluated exactly. To carry out this calculation, it is convenient to evaluate~$\frac{\partial I}{\partial t}$ first because this differentation removes the pole at~$s=0$. Then, we combine the integrals above and below the negative $x$-axis and remove the radicals by introducing~$p=\sqrt{s}$. The result reads

\begin{equation}
\label{I_prime}
\frac{\partial I}{\partial t} = \frac{\alpha v_0}{2\pi D_h\sqrt{D_ht}} \int_{0}^{+\infty}\frac{p^2\cos\frac{p|x|}{\sqrt{D_ht}}- \frac{\alpha v_0p}{4D_h}\sqrt{\frac{t}{D_h}}\sin\frac{p|x|}{\sqrt{D_ht}}}{p^2 + \frac{\alpha^2v_0^2}{16D_h^3}t} e^{-p^2} dp.
\end{equation}

This integral can be evaluated directly using the following formulas from Ref.~\cite{gradshteyn:tables}:

\begin{equation}
\begin{aligned}
& \int_0^{+\infty} e^{-bx^2}\cos(ax)dx = \frac{1}{2}\sqrt{\frac{\pi}{b}}e^{-\frac{a^2}{4b}},\\
& \int_0^{+\infty} e^{-bx^2}\sin(ax)\frac{xdx}{x^2+\gamma^2} = -\frac{\pi}{4}e^{b\gamma^2}\left[2\sinh(a\gamma) + e^{-a\gamma}\erf\left(\gamma\sqrt{b}-\frac{a}{2\sqrt{b}}\right) - e^{a\gamma}\erf\left(\gamma\sqrt{b}+\frac{a}{2\sqrt{b}}\right)\right],\\
& \int_0^{+\infty} e^{-bx^2}\cos(ax)\frac{dx}{x^2+\gamma^2} = \frac{\pi}{4\gamma}e^{b\gamma^2}\left[2\cosh(a\gamma) - e^{-a\gamma}\erf\left(\gamma\sqrt{b}-\frac{a}{2\sqrt{b}}\right) - e^{a\gamma}\erf\left(\gamma\sqrt{b}+\frac{a}{2\sqrt{b}}\right)\right].\\
\end{aligned}
\label{table_integrals}
\end{equation}

\noindent where~$\erf(x)$ is the error function~\cite{gradshteyn:tables}.

To obtain~$I$, we integrate~$\frac{\partial I}{\partial t}$ over~$t$; $I$ at~$t=0$ is determined from Eq.~(\ref{u_poles_symmetric}) and the fact that~$u(0,x)=1$. After judicious integration by parts, we find that

\begin{equation}
\label{u_symmetric_solution}
u(t,x) = e^{\frac{v_0^2\alpha^2}{16D_h^3}t-\frac{v_0\alpha}{4D_h^2}|x|}\erfc\left(\frac{|x|}{2\sqrt{D_ht}}-\frac{\alpha v_0}{4D_h}\sqrt{\frac{t}{D_h}}\right) + \erf\frac{|x|}{2\sqrt{D_ht}}, 
\end{equation}

\noindent where~$\erfc(x)$ is the complementary error function~\cite{gradshteyn:tables}. The solution for~$h(t,x)$ is obtained by changing variables back from~$u$ to~$h$:

\begin{equation}
h(t,x)_{\mathrm{bulge}} = v_0\left(1 + \frac{\alpha^2}{8D_h^2}\right)t-\frac{\alpha}{2D_h}|x| + \frac{2D_h}{v_0}\ln\left[\erfc\left(\frac{|x|}{2\sqrt{D_ht}}-\frac{\alpha v_0}{4D_h}\sqrt{\frac{t}{D_h}}\right) + e^{-\frac{v_0^2\alpha^2}{16D_h^3}t+\frac{v_0\alpha}{4D_h^2}|x|}\erf\left(\frac{|x|}{2\sqrt{D_ht}}\right)\right]. 
\label{bulge_symmetric_solution}
\end{equation}

\noindent For a fixed~$x$ and large~$t$, the last term in Eq.~(\ref{bulge_symmetric_solution}) approaches a constant, so the long-time limit is given by the first two terms, which specify the height and shape of the bulge.

\subsection*{Dip shape at a stationary out-flow boundary,~$\boldsymbol{f^*}=\frac{1}{2}$}
The analysis of the dip shape is essentially the same as above. Indeed, the substitution of~$f(t,x)=\theta(-x)$ into Eq.~(\ref{h_equation}) yields

\begin{equation}
\label{symmetric_dip_formulation}
\frac{\partial h}{\partial t} = v_0 + \frac{v_0}{2}\left(\frac{\partial h}{\partial x}\right)^2 + D_h \frac{\partial^2 h}{\partial x^2} - \alpha\delta(x),
\end{equation}

\noindent which is identical to Eq.~(\ref{symmetric_bulge_formulation}), but with the opposite sign in front of~$\alpha$. Therefore, the solution of Eq.~(\ref{symmetric_dip_formulation}) is obtained by replacing~$\alpha$ by~$-\alpha$ in Eq.~(\ref{bulge_symmetric_solution}):

\begin{equation}
\label{symmetric_dip_solution}
h(t,x)_{\mathrm{dip}} = v_0t + \frac{2D_h}{v_0}\ln\left[e^{\frac{v_0^2\alpha^2}{16D_h^3}t+\frac{v_0\alpha}{4D_h^2}|x|}\erfc\left(\frac{|x|}{2\sqrt{D_ht}}+\frac{\alpha v_0}{4D_h}\sqrt{\frac{t}{D_h}}\right) + \erf\frac{|x|}{2\sqrt{D_ht}}\right].
\end{equation}

\noindent Note that we stated this result in a slightly different from compared to Eq.~(\ref{bulge_symmetric_solution}) because the leading behavior in the long-time limit is now given by different terms; instead of approaching a constant, the complementary error function now tends to zero and nearly cancels the exponential term in front of it. For large~$t$ and finite~$x$, Eq.~(\ref{symmetric_dip_solution}) can be approximated by

\begin{equation}
\label{symmetric_dip_asymptotic}
h(t,x)_{\mathrm{dip}} \approx v_0t + \frac{2D_h}{v_0}\ln\left(\frac{\frac{8D_h^2}{\sqrt{\pi}\alpha v_0} + |x|}{2\sqrt{D_ht}}\right).
\end{equation}

\noindent Thus, the shape of a dip is a curvilinear angle made by two logarithmic curves rather than a regular angle made by two straight lines as we found in the case of a bulge; see Fig.4 in the main text. The depth of the dip,~$v_0t-h(t,0)$, increases logarithmically in time as~$\frac{D_h}{v_0}\ln\left(\frac{\pi\alpha^2v_0^2t}{16D_h^3}\right)$ and the width of the dip increases as~$2\sqrt{D_ht}$. Because both the depth and the width grow much slower for a dip than for a bulge, we expect that the front shape is largely dominated by the locations of the bulges. Consistent with this expectation, our simulation revealed only modest dips that do not extend appreciably beyond the width of the corresponding out-flow boundaries. This observation holds regardless of the value of~$f^*$.

\subsection*{The shape of a moving bulge,~$\boldsymbol{f^*}\ne\frac{1}{2}$}
We now relax the assumption that~$f^*=\frac{1}{2}$ and derive the asymptotic shape of a moving bulge. The calculation largely proceeds through the same steps as before starting with the following equation for~$h(t,x)$:

\begin{equation}
\label{moving_bulge_formulation}
\frac{\partial h}{\partial t} = v_0 + \frac{v_0}{2}\left(\frac{\partial h}{\partial x}\right)^2 + D_h \frac{\partial^2 h}{\partial x^2} + \alpha\delta(x-v_{\parallel}t).
\end{equation}

\noindent The change of variables from~$h$ to~$u$ according Eq.~(\ref{u}) results in

\begin{equation}
\label{u_moving_bulge}
\frac{\partial u}{\partial t} =D_h \frac{\partial^2 u}{\partial x^2} + \frac{v_0\alpha}{2D_h}u\delta(x-v_{\parallel}t).
\end{equation}

The next step is to shift into a reference frame moving with velocity~$v_{\parallel}$ by defining a new spatial variable~$\mathfrak{z}=x-v_{\parallel}t$. In terms of~$\mathfrak{z}$, the equation for~$u$ takes the following form

\begin{equation}
\label{u_moving_bulge}
\frac{\partial u}{\partial t} = D_h \frac{\partial^2 u}{\partial \mathfrak{z}^2} + v_{\parallel} \frac{\partial u}{\partial \mathfrak{z}}  + \frac{v_0\alpha}{2D_h}u\delta(\mathfrak{z}).
\end{equation}

We proceed by performing the Laplace transform in time, which yields

\begin{equation}
\label{u_moving_bulge}
D_h \frac{\partial^2 u}{\partial \mathfrak{z}^2} + v_{\parallel} \frac{\partial u}{\partial \mathfrak{z}} - su + \frac{v_0\alpha}{2D_h}u\delta(\mathfrak{z})  = -1,
\end{equation}

\noindent where we used the initial condition~$u(0,x)=1$. This linear equation can be solved for~$\mathfrak{z}>0$ and~$\mathfrak{z}<0$ separately under the constraints that~$\lim_{\mathfrak{z}\to-\infty}u=\lim_{\mathfrak{z}\to+\infty}u=0$. These two solutions are then matched by imposing continuity at~$\mathfrak{z}=0$ and the following condition due to the term with the delta function

\begin{equation}
\lim_{\mathfrak{z}\to+0}\frac{\partial u}{\partial \mathfrak{z}} - \lim_{\mathfrak{z}\to-0}\frac{\partial u}{\partial \mathfrak{z}} = -\frac{v_0\alpha}{2D^2_h}\lim_{\mathfrak{z}\to0}u.
\end{equation}

\noindent The result reads

\begin{equation}
\label{u_moving_bulge_solution_Laplace}
u = \frac{1}{s} \left[1 + \frac{\frac{\alpha v_0}{4D_h^2}}{\sqrt{\frac{s}{D_h}+ \frac{v_{\parallel}^2}{4D_h^2}} - \frac{\alpha v_0}{4D_h^2} }\left(\theta(\mathfrak{z})e^{-\left(\frac{v_{\parallel}}{2D_h}+\sqrt{\frac{s}{D_h}+ \frac{v_{\parallel}^2}{4D_h^2}}\right)\mathfrak{z}} + \theta(-\mathfrak{z})e^{-\left(\frac{v_{\parallel}}{2D_h}-\sqrt{\frac{s}{D_h}+ \frac{v_{\parallel}^2}{4D_h^2}}\right)\mathfrak{z}} \right)\right].
\end{equation}

As before,~$u(s,\mathfrak{z})$ has two poles and a branching point. The long-time behavior is determined by the singularity with the largest real part, which is either the pole at~$s=\frac{\alpha^2v_0^2}{16D_h^3}-\frac{v_{\parallel}^2}{4D_h}$, when~$\frac{\alpha^2v_0^2}{16D_h^3}-\frac{v_{\parallel}^2}{4D_h}>0$, or the pole at~$s=0$ otherwise. Let us consider the former possibility first, which occurs when~$|v_{\parallel}|<\frac{\alpha v_0}{2D_h}$.

Upon evaluating the contribution of the dominant pole to the inverse Laplace transform and changing from~$u$ to~$h$, we obtain that

\begin{equation}
\label{moving_bulge_solution}
h(t,x)_{\mathrm{bulge}} = \left(1 + \frac{\alpha^2}{8D_h^2} - \frac{v_{\parallel}^2}{2v_0^2} \right)v_0 t - \frac{\alpha}{2D_h}|x-v_{\parallel}t| - \frac{v_{\parallel}}{v_0}(x-v_{\parallel}t).
\end{equation}

\noindent Compared to a non-translating bulge for~$f^*=\frac{1}{2}$, a moving bulge exhibits a slower increase in height and an asymmetry between the leading and the trailing slopes. The leading slope is steeper while the trailing slope is shallower. 

As expected, the peak of the bulge, which corresponds to the location of the in-flow boundary, moves with~$v_{\parallel}$ relative to the $x$-axis. The spatial extent of the bulge is given by the locations where the sides of the bulge intersect the unperturbed flat front. The velocities of these bulge ends~(or bulge feet) can be easily determined from Eq.~(\ref{moving_bulge_solution}) by setting~$h$ to~$v_0t$. The results are

\begin{equation}
\label{v+v-}
\begin{aligned}
& v_{+} = \frac{v_{\parallel}}{2} + \frac{\alpha v_0}{4D_h},\\
& v_{-} = \frac{v_{\parallel}}{2} - \frac{\alpha v_0}{4D_h},\\
\end{aligned}
\end{equation}  

\noindent where~$+$ refers to the end on the right of the bulge and~$-$ to the end on the left of the bulge.

Note that one of these velocities as well as the slope of the trailing side of the bulge approach~$0$ as the~$|v_{\parallel}|$ increases towards~$\frac{\alpha v_0}{2D_h}$. This limit corresponds to the exclusion of the less chiral strain. Indeed, the in-flow boundary advances with velocity~$v_{\parallel}$ and a sharp slope at the leading edge. The shape of the leading edge does not change with time because the velocity of the corresponding bulge end also equals~$v_{\parallel}$. The velocity of the trailing end, on the other hand, is~$0$, so it remains stationary. As a result, the more chiral strain continually gains more and more territory in the direction of its chirality. Consistent with this picture, Eq.~(\ref{moving_bulge_solution}) also predicts that the slope of the trailing edge is~$0$ and the height of the bulge does not increase with time.

Once~$|v_{\parallel}|>\frac{\alpha v_0}{2D_h}$, the pole at~$s=0$ becomes the dominant singularity, and the asymptotic solution for~$h$ changes to

\begin{equation}
\label{moving_bulge_s_0}
h(t,x) = v_0t + \frac{2D_h}{v_0}\ln\left(1 + \frac{1}{\frac{2D_h v_{\parallel}}{\alpha v_0}-1}\left[\theta(x-v_{\parallel}t)e^{-\frac{v_{\parallel}}{D_h}(x-v_{\parallel}t)} + \theta(-x+v_{\parallel}t)\right]\right),
\end{equation}

\noindent where we assumed that~$v_{\parallel}>0$ to avoid cumbersome notation. The corresponding result for negative~$v_{\parallel}$ can obtained by applying a mirror symmetry.

The behavior of the bulge described by Eq.~(\ref{moving_bulge_s_0}) is completely analogous to the dynamics in the limit of~$v_{\parallel}$ approaching~$\frac{\alpha v_0}{2D_h}$ from below that we just discussed. In particular, the leading edge advances with velocity~$v_{\parallel}$ and has a time-invariant, in this case exponential, shape. The trailing edge has a zero slope, but a slightly higher height compared to the unperturbed flat front away from the in-flow boundary. Therefore, we conclude that, when~$|v_{\parallel}|\ge \frac{\alpha v_0}{2D_h}$, the more chiral strain invades the less chiral strain, but the less chiral strain does not invade the more chiral strain.

\section{Natural selection due to the influence of front shape on the dynamics of~$\boldsymbol{f(t,x)}$}

This section explains how bulge growth leads to natural selection via the mechanism illustrated in Fig.5 in the main text. The key idea is that the motion of an out-flow boundary changes once it comes in contact with the bulge. Here, we only consider~$v_{\parallel}<\frac{\alpha v_0}{2D_h}$ because the mechanism of selection for~$|v_{\parallel}|\ge \frac{\alpha v_0}{2D_h}$ has been explained in the preceeding section.

\subsection*{Boundary motion in the presence of bulges}
For concreteness, let us assume that~$v_{\parallel}>0$ and that the bulge reaches the out-flow boundary from the left, as shown in Figs.5 and 6 in the main text. The analysis of the other cases is completely analogous. 

The first thing to observe is that the contact of an out-flow boundary with a bulge is inevitable. An out-flow boundary located within a flat region of the front moves with velocity~$v_{\parallel}$ specified by Eq.~(\ref{v_parallel}). This velocity is opposite to the velocity of the left end of the bulge,~$v_{-}$, and is smaller than the velocity of the right end of the bulge,~$v_{+}$; see Eq.~(\ref{v+v-}). Thus, an out-flow boundary and one of the nearest bulges always meet.

The second thing to observe is that an out-flow boundary on the slope of a bulge catches up with the bulge end. To demonstrate this, we need to account for the effect of front tilt on boundary motion. Let us consider a region, such as the slope of the bulge, where~$\frac{\partial h}{\partial x}=\varphi=\mathrm{const}$. Equation~(\ref{f_equation}) then takes the following form:

\begin{equation}
\label{f_equation_slope}
\frac{\partial f }{\partial t}  = D_f\frac{\partial^2 f}{\partial x^2} + \beta (f^* - f) \frac{\partial f}{\partial x} +  v_0 \varphi \frac{\partial f}{\partial x}.
\end{equation}

\noindent Thus, the coupling between~$h$ and~$f$ leads to an effective advection with velocity~$-v_0\varphi$, which modifies our expression for the boundary velocity in Eq.~(\ref{v_parallel}) as follows 

\begin{equation}
\label{boundary_motion}
v_{b} = v_{\parallel} - v_0\varphi,
\end{equation}

\noindent where we used~$v_b$ to denote the boundary velocity within a titled region of the front and~$v_{\parallel}$ to denote the boundary velocity within a flat region of the front. Upon evaluating~$\varphi$ from Eq.~(\ref{moving_bulge_solution}), we find that

\begin{equation}
\label{boundary_on_bulge}
v_{b} = 2v_{\parallel} + \frac{\alpha v_0}{2D_h}  >  v_{+},
\end{equation}

\noindent i.e. an out-flow boundary on the slope of the bulge moves faster than the bulge end.

From the two observations above, we conclude that an out-flow boundary must localize at the end of the bulge. Indeed, the boundary can neither escape in the flat region ahead of the front nor fall behind on the slope of the bulge. While our asymptotic solution predicts a discontinous change in~$\frac{\partial h}{\partial x}$ when the bulge and the flat front meet, it is clear that the slope of the front changes continuously. The inequality in Eq.~(\ref{boundary_on_bulge}) ensures that there is a value of~$\frac{\partial h}{\partial x}$ in this transition region such that Eq.~(\ref{boundary_motion}) is satisfied.

\subsection*{The origin of selection}
After an out-flow boundary is trapped at the bulge end, it moves with velocity~$v_{+}$ rather than~$v_{\parallel}$. This leads to a change in the relative abundance of the two strains, i.e. to natural selection. The change in the strain fractions ceases only when the out-flow boundary is locked between the opposite slopes of the two nearest bulges. In the simulation shown in Fig.5 in the main text, the symmetry due to~$f^*=1/2$ ensures that the out-flow boundaries are exactly in the middle between the two bulges. Therefore, selection stops when the strains reach equal fractions. The dynamics for~$f^*\ne1/2$ are discussed next.

\subsection*{Negative frequency-dependent selection and the derivation of~$\boldsymbol{\bar{f}_{\mathrm{eq}}}$}
Since natural selection is mediated by the motion of out-flow boundaries, the steady state is reached when the out-flow boundaries are no longer moving relative to each other. This relative motion ceases when all out-flow boundaries are in contact with both of the nearest bulges as illustrated in Fig.6 in the main text. In this coexistence phase, the relative abundance of the strains can be quantified by~$\bar{f}(t)$, the spatial average of~$f(t,x)$. 

The value of~$\bar{f}_{\mathrm{eq}}$ in equilibrium can be computed using the following geometrical argument; see Fig.6 in the main text. For small~$\frac{\partial h}{\partial x}$, the height of the bulge equals the product of its slope and the horizontal extent of its side. This product must be the same for both sides of the bulge because the height of the bulge has a unique value. Therefore, the ratio of the side lengths is equal to the inverse ratio of the side slopes. Since the ratio of the slopes is the same for all bulges, so must be the ratio of the sizes of the domains occupied by each of the strains. The latter quantity is nothing but~$\bar{f}_{\mathrm{eq}}/(1-\bar{f}_{\mathrm{eq}})$, so we conclude that

\begin{equation}
\label{coexistence_geometry}
\frac{\bar{f}_{\mathrm{eq}}}{1-\bar{f}_{\mathrm{eq}}} = \frac{\left.\frac{\partial h}{\partial x}\right|_{\mathrm{side\;\;occupied\;\;by\;\;strain\;\;2}}}{\left.\frac{\partial h}{\partial x}\right|_{\mathrm{side\;\;occupied\;\;by\;\;strain\;\;1}}}.
\end{equation}

The value of~$\bar{f}_{\mathrm{eq}}$ can then be obtained by using Eq.~(\ref{moving_bulge_solution}) to determine bulge slopes and Eq.~(\ref{v_parallel}) to determine~$v_{\parallel}$. The result reads

\begin{equation}
\label{f_bar}
\bar{f}_{\mathrm{eq}} = \frac{1}{2} + \frac{\beta D_{h}}{\alpha v_0}\left(f^*- \frac{1}{2}\right).
\end{equation}

For~$|f^*-\frac{1}{2}| < \frac{\alpha v_0}{2D_h\beta}$, Eq.~(\ref{f_bar}) predicts that~$\bar{f}_{\mathrm{eq}}\in(0,1)$, i.e. both strains are present at steady state. When this inequality is violated,~$|v_{\parallel}|\ge \frac{\alpha v_0}{2D_h}$, and one of the strains goes extinct as we discussed above. This transition occurs at~$f_{c}^* = \frac{1}{2} \pm \frac{\alpha v_0}{2D_h\beta}$, which could lie both within and outside~$[0,1]$. In the former case, the competition of the non-chiral strain vs. a chiral strain~($f^*=0$ or~$f^*=1$) always results in the extinction of the less chiral strain. In the latter case, a chiral and a non-chiral strains coexist. 

To avoid possible confusion, we emphasize that, unlike the value of~$\bar{f}$, the value of~$f^*$ is not constrained to lie between~$0$ and~$1$; see Eq.~(\ref{f*}). The values of~$f^*$ outside~$[0,1]$ correspond to strains that have different magnitudes of chirality, but are chiral in the same direction~(i.e.~$A^{(1)}$ and~$A^{(2)}$ have the same sign). In particular, two strains with the same chirality correspond to~$\alpha=0$,~$\beta=0$,~$f^*=\infty$, and~$\beta f^*=\mathrm{finite}$.


\section{Off-lattice simulations}

Here we describe the simulations that were performed without introducing a lattice of sub-populations and, therefore, without breaking the rotational invariance of the space. These simulations provide an important confirmation of our theoretical results because lattice effects are known to create many artifacts especially in the presence of chirality~\cite{LauraTucker:chiral_lattice_effects}.

In off-lattice simulations, each cell~$i$ has position~$(x_i, y_i)$ which are continuous variables without an underlying lattice. Cell growth and movement depends on the arrangement of other cells through the effective population density,~$C(x, y)$ and its gradient~$\boldsymbol{\nabla} C(x, y)$.
The effective population density sensed by cell~$i$ is given by
\begin{equation}
\label{off_lattice_concentration}
C(x_i,y_i)=\sum\limits_{j} \frac{1}{N} e^{-(x_i-x_j)^2-(y_i-y_j)^2},
\end{equation}
where the sum is over the entire population of cells, and~$N$ is the carrying capacity which normalizes~$C$ to be order~$1$ in the bulk of the population. Note that the Gaussian kernel is rotationally and translationally symmetric. The gradient of the population density is computed from its definition and is given by
\begin{equation}
\label{off_lattice_gradient}
\boldsymbol{\nabla} C (x,y)=  -2\sum\limits_{j} \frac{1}{N} e^{-(x-x_j)^2-(y-y_j)^2} \left[(x-x_j) \hat{\textsf{x}} +(y-y_j) \hat{\textsf{y}} \right].
\end{equation}

A cell~$i$ at~$(x_i, y_i)$ gives birth to a daughter cell at the same position with probability~$g_i(1- C(x_i,y_i))$, where~$g_i$ is the growth rate of the cell~$i$.  If the region is overcrowded, the growth probability is negative and the cell dies with the negative of this growth probability.

A cell at~$(x_i, y_i)$ jumps with a probability proportional to the population density~$C(x_i,y_i)$. This cooperative migration ensures that the wave is pushed. The jump size is fixed to a value~$\mu$. Since cell motion is chiral, the direction of the jump is biased by the gradient in the concentration. The angle of the jump of cell~$i$ with chirality~$A_i$ is termed~$\theta_i$. This angle is chosen  from a distribution with mean~$\phi_i$ such that
\begin{equation}
\label{off_lattice_bias}
\tan{\phi_i}= - \frac{ \boldsymbol{\nabla} C(x_i,y_i) \textbf{.}\hat{\textsf{x}} }{\boldsymbol{\nabla} C(x_i,y_i) \textbf{.} \hat{\textsf{y}}}
\end{equation}

where we ensure that~$\phi_i$ is in the correct quadrant accounting for the sign of the chirality. The orientation of~$\phi$ relative to the gradient was~$90^{\circ}$ counterclockwise for positive chirality and~$90^{\circ}$ clockwise for negative chirality. Although the mean direction of the jump could be at an arbitrary angle with respect to the gradient in the most the most general model, we  restrict ourselves to this case which was sufficient to explore the effects of chirality. The angle of the jump,~$\theta_i$ is then chosen from the wrapped normal distribution

\begin{equation}
\label{off_lattice_theta}
P(\theta_i)= \sqrt{\frac{|\boldsymbol{\nabla} C(x_i,y_i)| |A_i|  }{\pi^2 }} \sum\limits_{k= -\infty}^{k=\infty}  \textsf{exp}\left[- \frac{   \left( \theta_i -\phi_i +2 \pi k \right)^2  }{\pi }|\boldsymbol{\nabla} C(x_i,y_i)| |A_i|  \right] ,
\end{equation}

where~$\theta_i$ lies in ~$[0,2 \pi)$.
The variance of the normal distribution is inversely proportional to magnitude of the chirality of the cell and the population gradient. This ensures that the angle of the jump is sharply peaked if the cell is very chiral and the gradient is strong. When either the gradient or the chirality is weak, the distribution tends to a uniform distribution which leads to unbiased diffusion.  To suppress migration into crowded areas, a jump to position~$(x_d,y_d)$ is rejected with probability~$\left(1-C(x_d,y_d)\right)$. This freezes dynamics in the colony bulk like in the experiments.

In simulations, the cells grow in a semi-infinite box extending in the $+y$ direction, with finite width $W$ and periodic boundary conditions in the $x$ direction. We use synchronous updates: first the effective population density and its gradient are computed for every cell, then the cells are allowed to grow or die, and finally the migration step is performed. To speed up simulations, we only simulate cells near the front, as the dynamics in the colony bulk are frozen and activity in the bulk has no effect on the front dynamics.


\newpage

\section{Additional simulation results}


\begin{figure}
\begin{center}
\includegraphics[width=\linewidth]{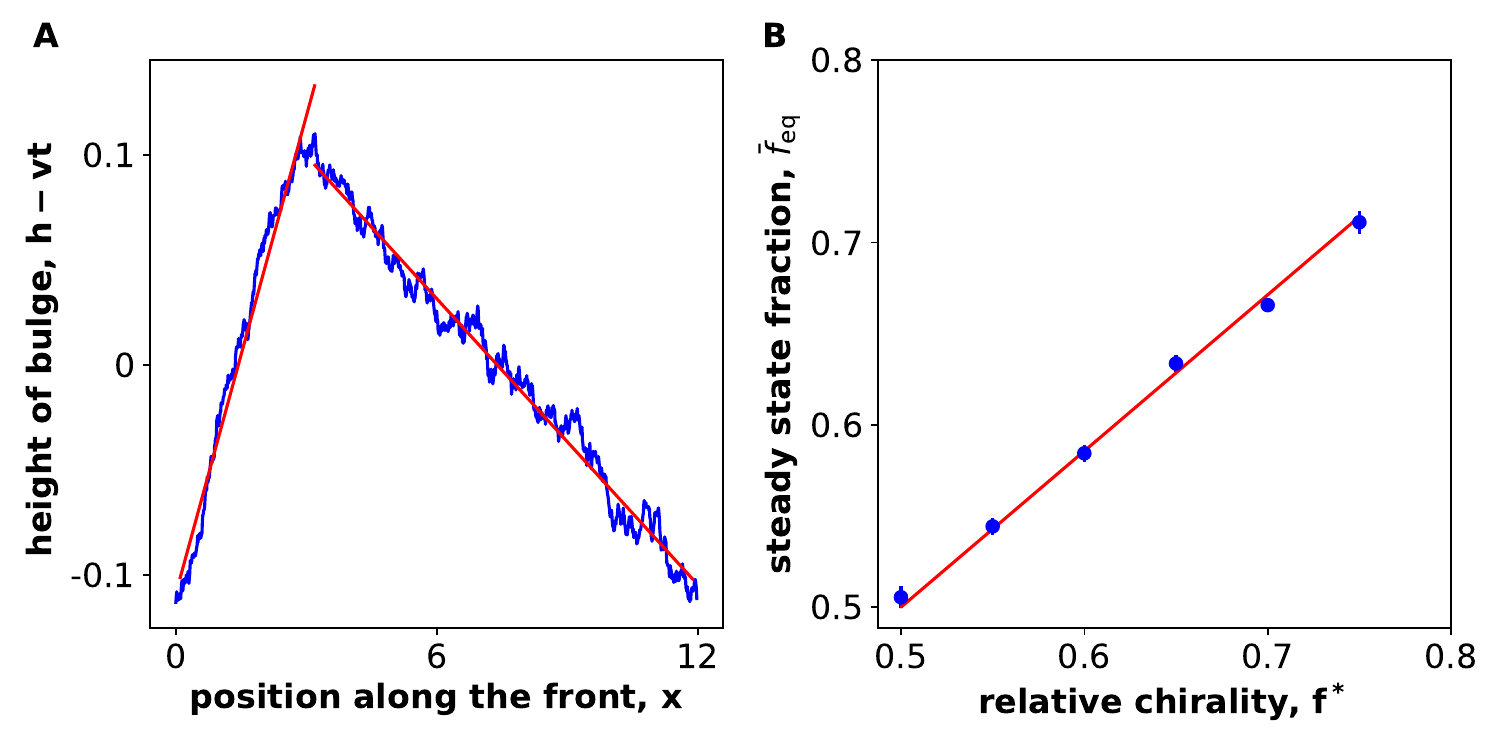}
\caption{{\bf Quantitative test of Eq.~(5), the relationship between~$\bar{f}_{\mathrm{eq}}$ and~$f^*$.} \textbf{(A)} shows the shape~(shape preserving spline) of an asymmetric bulge from a simulation with~$f^*=0.75$. The red lines are the best fit of the two slopes of the bulge. From these slopes, we obtained~$\frac{\alpha}{D_h}=0.080$ and~$\frac{\beta}{v_0}=0.068$ using Eq.~(S55); the values of these parameters are averages over all runs with~$f^* \neq 0.5$. \textbf{(B)}~shows~$\bar{f}_{\mathrm{eq}}$ from simulations~(dots) and the theoretical prediction~(line) from Eq.~(5) in the main text and the estimated values of the parameters. The predicted slope equals~$0.856$ and is quite close to~$0.82 \pm 0.02$, which is the slope obtained by ordinary least squares regression~(not shown). The root mean square deviation between the theory and the simulations is $0.01$. Here,~$m_0=m_s=m_b=m_d=0$,~$g=0.1$,~$m_l+m_r=0.01$ for both strains. Simulations started from two separate domains on a lattice of~$1200 \textsf{x} 14000$ sites with~$N=800$. We ensured that the simulations reached steady state by starting runs from an initial fractions of $0.25$,~$0.5$ and ~$0.75$. Error bars~(s.e.m) were estimated from a set of~18 runs with~6 starting from each initial fraction.
}
\label{S1_Fig}
\label{fig:S1}
\end{center}  
\end{figure}

\begin{figure}
\begin{center}
\includegraphics[width=8.7cm]{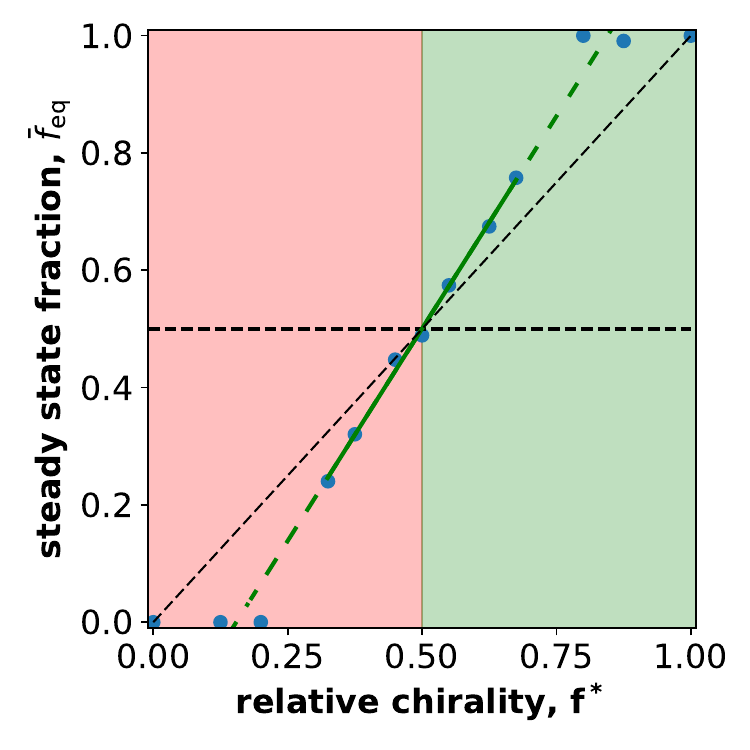}
\caption{{\bf The size of the coexistence region depends on model parameters.} This figure is the same as Fig.~7B, but for different model parameters. In comparison with the figure in the main text, the coexistence region is smaller, and the slope of~$\bar{f}_{\mathrm{eq}}$ is steeper. The green line is the least squares fit to the simulation data~(dots); the slope is~$1.45$, and~$R^2=0.997$. The solid part of the line spans the data points with~$\bar{f}_{\mathrm{eq}}\in(0,1)$ that were used in the fit. The dashed part extends this dependence to the entire region of possible~$\bar{f}_{\mathrm{eq}}$. The black dashed line marks the unit slope. Here,~$m_0=0.01,m_s=m_b=m_d=0$,~$g=0.1$,~$m_l+m_r=0.1$ for both strains. Simulations started from well mixed conditions on a lattice of width~$300$ sites with~$N=200$. The simulation time was chosen to ensure that the same steady state was approached starting from initial conditions that are both above and below~$\bar{f}_{\mathrm{eq}}$. 
}
\label{S2_Fig}
\end{center}  
\end{figure}

\begin{figure}
\begin{center}
\includegraphics[width=\linewidth]{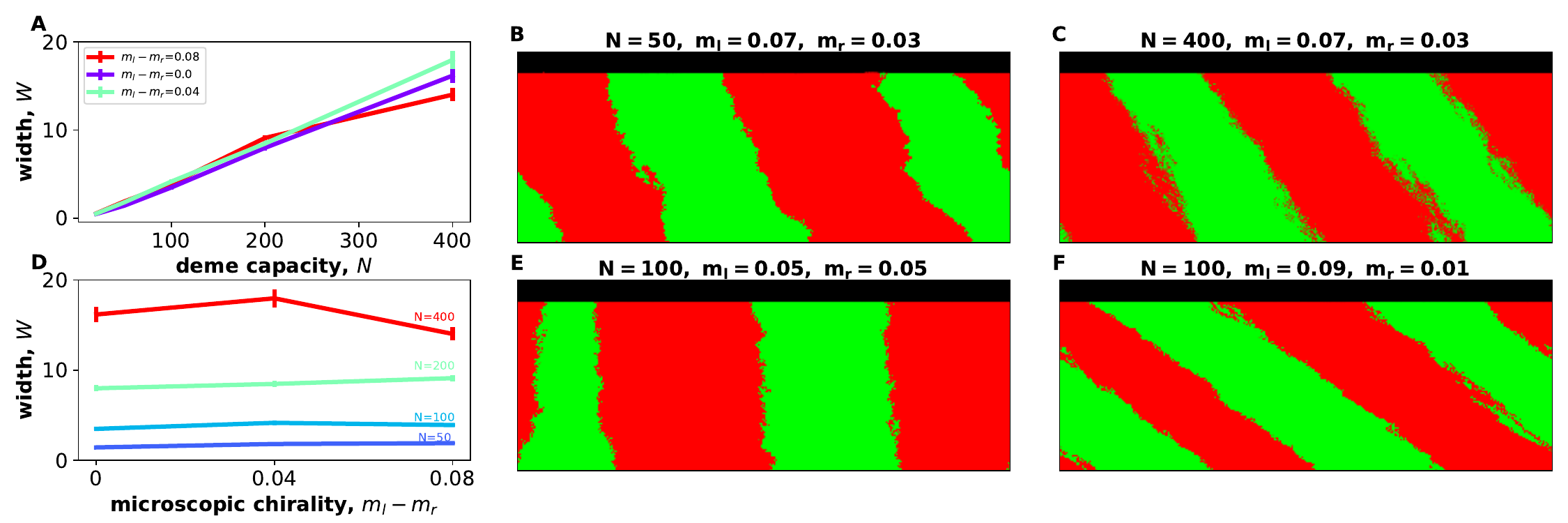}
\caption{{\bf Genetic drift, but not chirality control the boundary width between identical strains.} \textbf{(A)}~shows that domain boundaries between strains with equal chirality become wider for larger $N$~(weaker genetic drift); the dependence is approximately linear in agreement with Ref.~\protect{\cite{hallatschek:noisy_fisher}}. \textbf{(D)} In contrast, chirality has no detectable effect on the boundary width. The rest of the panels show the spatial patterns used to reach these conclusions. The boundary width was computed as the local heterozygosity,~$2\langle f(1-f)\rangle$, summed over the entire width of the simulation and averaged over~$y\in(2000, 4000)$. Simulations were performed on a lattice of~$1000 \textsf{x} 4000$ sites with~$m_0=m_s=m_b=m_d=0$,~$g=0.1$.
}
\label{S3_Fig}
\end{center} 
\end{figure}

\begin{figure}
\begin{center}
\includegraphics[width=\linewidth]{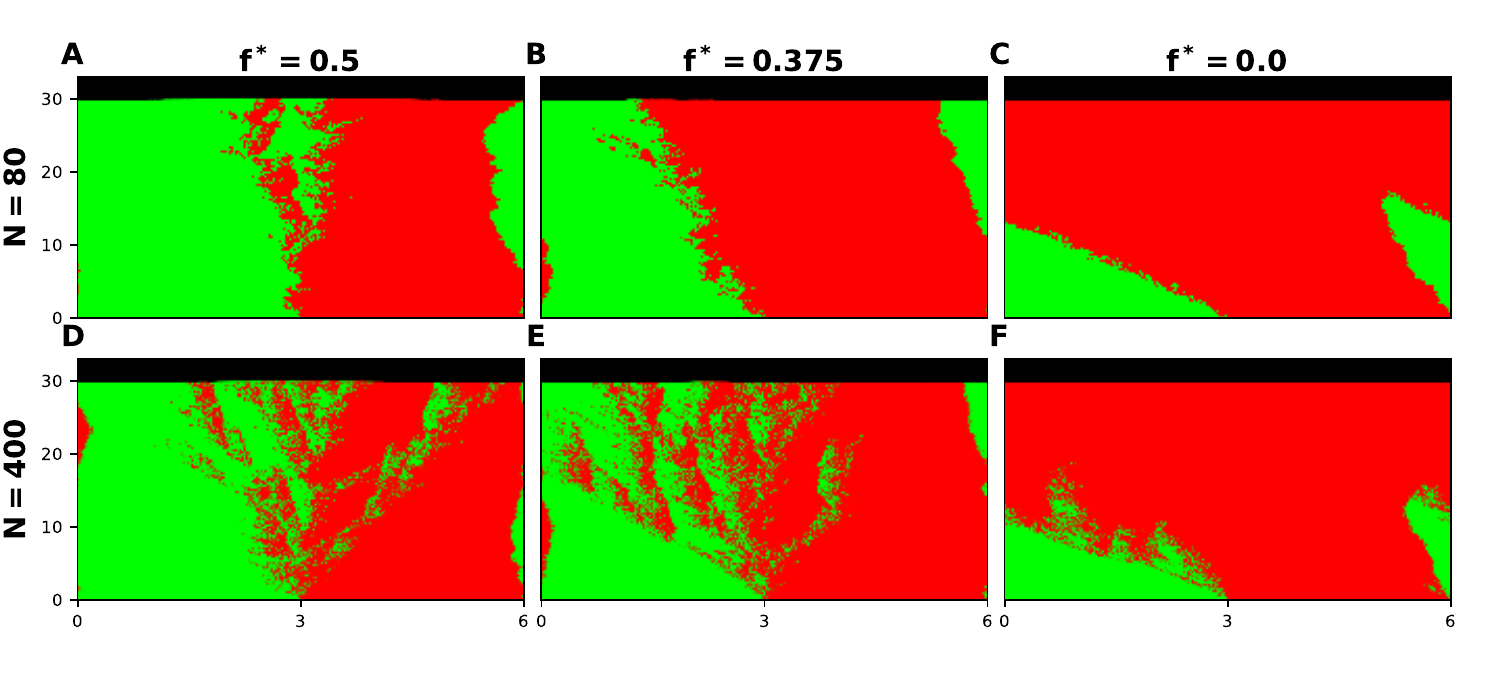}
\caption{{\bf Mixing transitions at different values of relative chirality.} \textbf{(A), (B), (C)}~show demixed phases for strong genetic drift and varying values of~$f^*$. \textbf{(D), (E)}~shows dissolution of a boundary and the establishment of the intermixed phase for weak genetic drift and two values of~$f^*$. \textbf{(F)}~shows the competition between a chiral and a non-chiral strain for weak genetic drift. The boundary is much wider than in~(C), but no intermixed phase is established because the non-chiral strain is outcompeted. Here,~$m_0=m_s=m_b=m_d=0$,~$g=0.1$,~$m_l+m_r=0.1$ for both strains. Simulations were carried out on a lattice of~$600 \textsf{x} 3000$ sites.
}
\label{S4_Fig}
\end{center}  
\end{figure}

\begin{figure}
\begin{center}
\includegraphics[width=\linewidth]{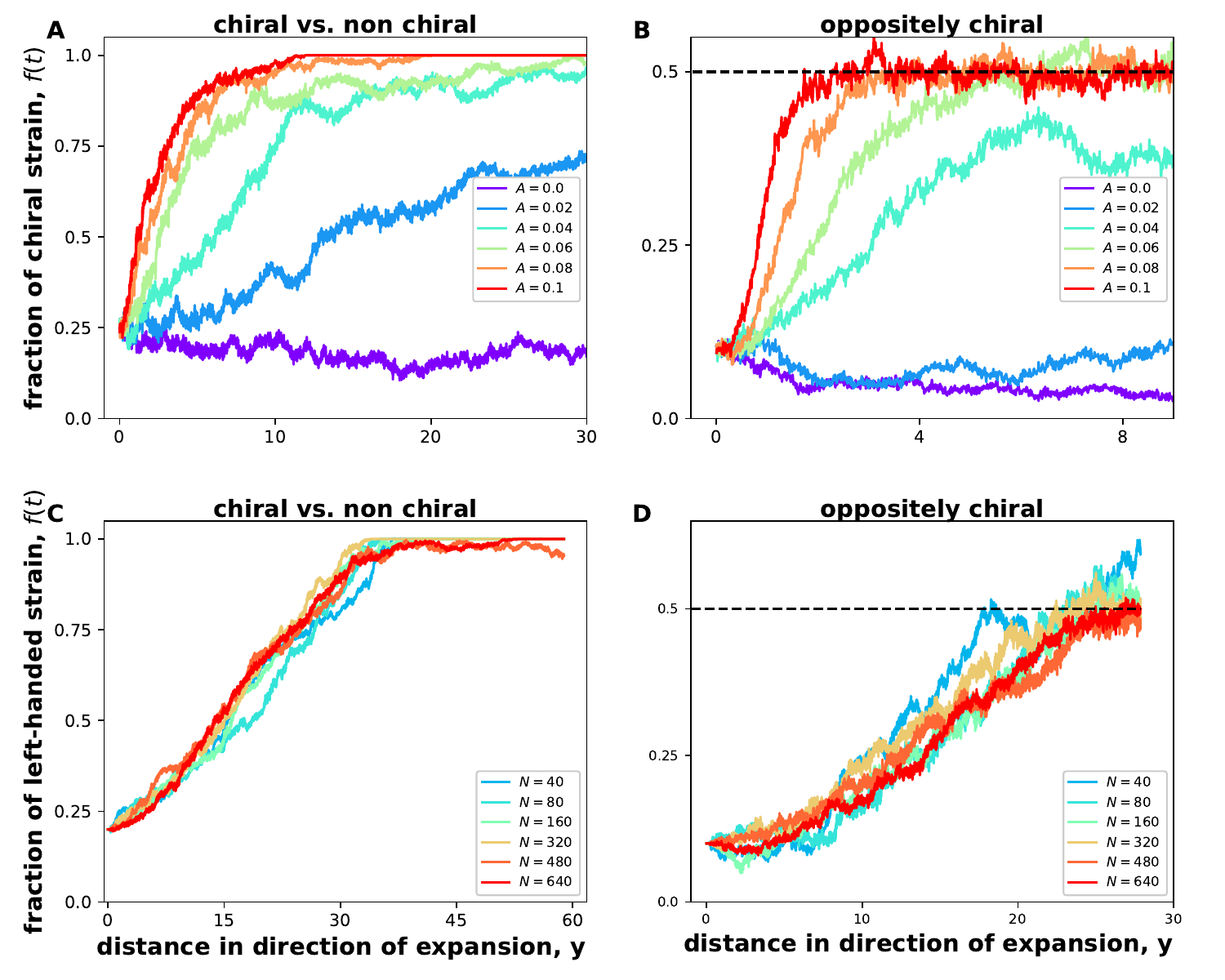}
\caption{{\bf Mixing transitions at different values of relative chirality.} \textbf{(A), (B), (C)}~show demixed phases for strong genetic drift and varying values of~$f^*$. \textbf{(D), (E)}~shows dissolution of a boundary and the establishment of the intermixed phase for weak genetic drift and two values of~$f^*$. \textbf{(F)}~shows the competition between a chiral and a non-chiral strain for weak genetic drift. The boundary is much wider than in~(C), but no intermixed phase is established because the non-chiral strain is outcompeted. Here,~$m_0=m_s=m_b=m_d=0$,~$g=0.1$,~$m_l+m_r=0.1$ for both strains. Simulations were carried out on a lattice of~$600 \textsf{x} 3000$ sites.
}
\label{S5_Fig}
\end{center} 
\end{figure}

\begin{figure}
\begin{center}
\includegraphics[width=\linewidth]{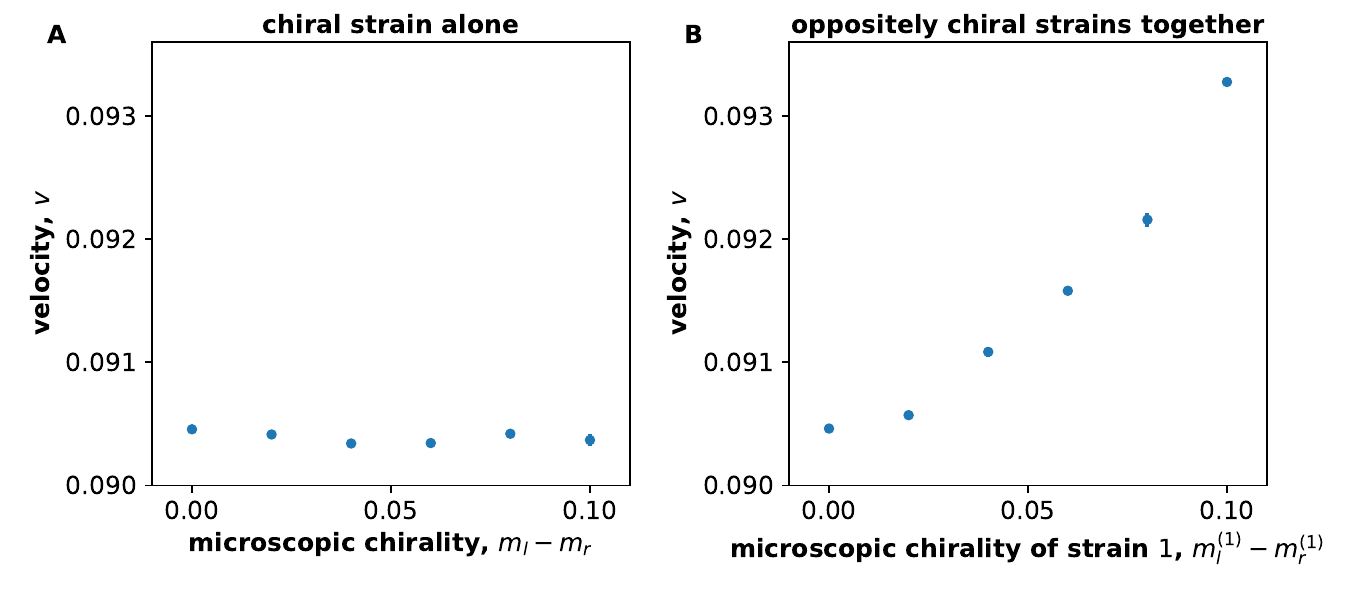}
\caption{{\bf Effects of chirality on the expansion velocity.} In panel \textbf{(A)}, we demonstrate that a change in chirality does not produce a change in the expansion velocity of a strain grown in isolation when~$m_l+m_r$ is kept fixed. \textbf{(B)}~shows that the expansion velocity increases, but only slightly, when two strains with opposite handedness expand together. Here,~$m_0=m_s=m_b=m_d=0$,~$g=0.1$,~$m_l+m_r=0.1$ for both strains. Simulations started from well-mixed conditions on a lattice of~$200 \textsf{x} 1000$ sites with~$N=200$. Error bars~(s.e.m) were estimated from~$4$ identical runs and velocity is measured in units of lattice spacing.
}
\label{S6_Fig}
\end{center} 
\end{figure}

\begin{figure}
\begin{center}
\includegraphics[width=\linewidth]{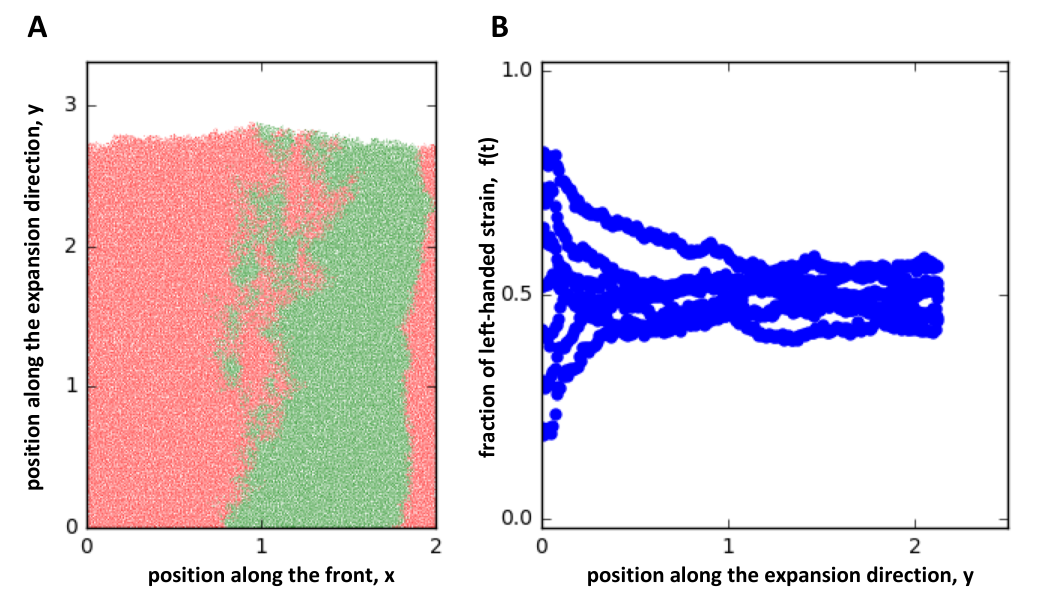}
\caption{{\bf Chirality affects competition in off-lattice simulations.} \textbf{(A)}~shows the emergence of a bulge between two oppositely chiral strains. \textbf{(B)}~shows the stabilizing selection between two oppositely chiral strains similar to Fig.~3C in the main text. The species fractions were averaged over~$6$ runs. Parameters were~$N=6$,~$A^{(1)}=10$,~$A^{(2)}=-10$,~$g^{(1)}=g^{(2)}=0.2$ for both figures, and~$W=200$,~$\mu=1.0$ in (A) and~$W=500$,~$\mu=0.5$ in (B). Distances were rescaled by a factor of~$100$, similar to simulations on the lattice in the text.
}
\label{S7_Fig}
\end{center}  
\end{figure}

\FloatBarrier

\bibliography{final_version_withSI_in_file}
\bibliographystyle{plos2015.bst}

\end{document}